\input harvmac
\def\journal#1&#2(#3){\unskip, \sl #1\ \bf #2 \rm(19#3) }
\def\andjournal#1&#2(#3){\sl #1~\bf #2 \rm (19#3) }

\def\ie{{\it i.e.}}
\def\eg{{\it e.g.}}

\def\frac#1#2{{#1\over#2}}

\def\half{\frac12}

\def\inbar{\,\vrule height1.5ex width.4pt depth0pt}
\def\IC{\relax\hbox{$\inbar\kern-.3em{\rm C}$}}
\def\IR{\relax{\rm I\kern-.18em R}}
\def\IP{\relax{\rm I\kern-.18em P}}
\def\IZ{\relax{\rm I\kern-.18em Z}}

%
%

%
\catcode`\@=11
\def\slash#1{\mathord{\mathpalette\c@ncel{#1}}}
\overfullrule=0pt

\def\JJ{{\cal J}}

\def\NN{{\cal N}}

\def\RR{{\cal R}}
\def\SS{{\cal S}}

\def\underrel#1\over#2{\mathrel{\mathop{\kern\z@#1}\limits_{#2}}}

\catcode`\@=12


%

\def \sinh{{\rm sinh}}
\def \cosh{{\rm cosh}}

\def\exp{{\rm exp}}


\rightline{RI-6-01, EFI-01-20}
\Title{
\rightline{hep-th/0106005}}
{\vbox{\centerline{Comments on D-branes in $AdS_3$}}}
\smallskip
\centerline{\it Amit Giveon${}^{1}$, David Kutasov${}^{2,3}$ and
Adam Schwimmer${}^{3}$}
\bigskip
\centerline{${}^1$Racah Institute of Physics, The Hebrew University}
\centerline{Jerusalem 91904, Israel}
\smallskip
\centerline{${}^2$Department of Physics, University of Chicago}
\centerline{5640 S. Ellis Av., Chicago, IL 60637, USA }
\smallskip
\centerline{${}^3$Department of Physics, Weizmann Institute of Science}
\centerline{Rehovot 76100, Israel}

\bigskip\bigskip\bigskip
\noindent
We study D-branes that preserve a diagonal $SL(2)$ affine Lie algebra
in string theory on $AdS_3$. We find three classes of solutions,
corresponding to the following representations of $SL(2)$:
(1) degenerate, finite dimensional representations with half integer
spin, (2) principal continuous series, (3) principal discrete series.
We solve the bootstrap equations for the vacuum wave functions and
discuss the corresponding open string spectrum. We argue that from
the point of view of the AdS/CFT correspondence, the above D-branes
introduce boundaries with conformal boundary conditions into the two
dimensional spacetime. Open string vertex operators correspond to
boundary perturbations. We also comment on the geometric interpretation
of the branes.

\vfill

\Date{5/01}

\newsec{Introduction}

\lref\gksch{A. Giveon and D. Kutasov,
``Notes on $AdS_3$,'' [hep-th/0106004].}
\lref\gks{A.~Giveon, D.~Kutasov and N.~Seiberg,
``Comments on string theory on AdS(3),''
Adv.\ Theor.\ Math.\ Phys.\  {\bf 2}, 733 (1998)
[hep-th/9806194].}
\lref\ks{D.~Kutasov and N.~Seiberg,
``More comments on string theory on AdS(3),''
JHEP {\bf 9904}, 008 (1999) [hep-th/9903219].}
\lref\maloog{J.~Maldacena and H.~Ooguri,
``Strings in AdS(3) and SL(2,R) WZW model. I,''
[hep-th/0001053].}
\lref\ags{R.~Argurio, A.~Giveon and A.~Shomer,
``Superstrings on AdS(3) and symmetric products,''
JHEP {\bf 0012}, 003 (2000) [hep-th/0009242].}
\lref\nw{C.~R.~Nappi and E.~Witten,
``A Closed, expanding universe in string theory,''
Phys.\ Lett.\ B {\bf 293}, 309 (1992) [hep-th/9206078].}
\lref\gm{P.~Ginsparg and G.~Moore,
``Lectures On 2-D Gravity And 2-D String Theory,''
[hep-th/9304011].}
\lref\ov{H.~Ooguri and C.~Vafa,
``Two-Dimensional Black Hole and Singularities of CY Manifolds,''
Nucl.\ Phys.\ B {\bf 463}, 55 (1996) [hep-th/9511164].}
\lref\gkp{
A.~Giveon, D.~Kutasov and O.~Pelc,
``Holography for non-critical superstrings,''
JHEP {\bf 9910}, 035 (1999)
[hep-th/9907178].}
\lref\gk{A.~Giveon and D.~Kutasov,
``Little string theory in a double scaling limit,''
JHEP {\bf 9910}, 034 (1999) [hep-th/9909110].}
\lref\dk{D.~Kutasov, ``Introduction to Little String Theory,''
2001 Trieste Lectures.}
\lref\gktwo{
A.~Giveon and D.~Kutasov,
``Comments on double scaled little string theory,''
JHEP {\bf 0001}, 023 (2000)
[hep-th/9911039].}
\lref\sahak{D.~Kutasov and D.~A.~Sahakyan,
``Comments on the thermodynamics of little string theory,''
JHEP {\bf 0102}, 021 (2001) [hep-th/0012258].}
\lref\kkk{V.~Kazakov, I.~K.~Kostov and D.~Kutasov,
``A matrix model for the two dimensional black hole,''
[hep-th/0101011].}
\lref\malnun{J.~M.~Maldacena and C.~Nunez,
``Towards the large N limit of pure N = 1 super Yang Mills,''
Phys.\ Rev.\ Lett.\  {\bf 86}, 588 (2001) [hep-th/0008001].}
\lref\egk{S.~Elitzur, A.~Giveon and D.~Kutasov,
``Branes and N = 1 duality in string theory,''
Phys.\ Lett.\ B {\bf 400}, 269 (1997)
[hep-th/9702014].}
\lref\efr{S.~Elitzur, A.~Forge and E.~Rabinovici,
``Some global aspects of string compactifications,''
Nucl.\ Phys.\ B {\bf 359}, 581 (1991).}
\lref\msw{G.~Mandal, A.~M.~Sengupta and S.~R.~Wadia,
``Classical solutions of two-dimensional string theory,''
Mod.\ Phys.\ Lett.\ A {\bf 6}, 1685 (1991).}
\lref\witten{E.~Witten,
``On string theory and black holes,''
Phys.\ Rev.\ D {\bf 44}, 314 (1991).}
\lref\PetropoulosNC{
P.~M.~Petropoulos,
``String theory on AdS(3): Some open questions,''
hep-th/9908189.
}
\lref\ArgurioIR{
R.~Argurio, A.~Giveon and A.~Shomer,
``String theory on AdS(3) and symmetric products,''
Fortsch.\ Phys.\  {\bf 49}, 409 (2001)
[hep-th/0012117].
}
\lref\ElitzurPQ{S.~Elitzur, A.~Giveon, D.~Kutasov, E.~Rabinovici and
G.~Sarkissian, ``D-branes in the background of NS fivebranes,''
JHEP {\bf 0008}, 046 (2000) [hep-th/0005052].}
\lref\adg{I.~Antoniadis, S.~Dimopoulos and A.~Giveon,
``Little string theory at a TeV,''
hep-th/0103033.}

In this paper we continue our study \gksch\ of string theory on $AdS_3$,
focusing on the physics of D-branes in this background. There are
a number of motivations for studying such D-branes (see also \gksch):
\item{(1)} The $SL(2,R)$ group manifold is one of the simplest
non-compact curved backgrounds in string theory. Since D-branes are
believed to be important for a microscopic understanding of string
theory, it seems useful to develop a better understanding of D-brane
dynamics in such backgrounds.

\item{(2)} In the context of the AdS/CFT correspondence, the $AdS_3$ case
is special in a number of ways: the relevant two dimensional conformal group
is infinite dimensional, and the theory can be studied beyond the supergravity
approximation, since it can be defined without turning on Ramond-Ramond
backgrounds (see \eg\
\refs{\gks,\ks,\PetropoulosNC,\maloog,\ArgurioIR} for some recent
discussions and additional references).
It is thus interesting to study D-branes in $AdS_3$ to address
various questions regarding the correspondence. For example, given a
string background of the form $AdS_3\times\NN$, where $\NN$ is a compact
manifold\foot{See \eg\ \ags\ for a construction of a large class of such
backgrounds.}, can one construct directly
the dual ``spacetime'' two dimensional CFT, perhaps by studying D-branes
in the bulk theory? Other (related) questions are what do
D-branes in $AdS_3$ correspond to in the spacetime CFT? What excitations
live on the branes and what is their role in the spacetime CFT?

\item{(3)} As a special case of (1) above, there are some interesting
backgrounds that are closely related to $SL(2,R)$; an understanding
of D-branes in $AdS_3$ would be very useful for studying D-branes in these
backgrounds. For example, Liouville theory and the Euclidean and Lorentzian
two dimensional black hole are cosets of the form $SL(2)/U(1)$
\refs{\efr,\msw,\witten}; Liouville corresponds to modding out by a Borel
subgroup, while the Euclidean (Lorentzian)
black hole is obtained by modding out by the timelike (spacelike) $U(1)$.
The cosmological model of \nw\ corresponds to the coset $[SL(2)\times SU(2)]
/U(1)^2$. All the above backgrounds are of interest in the context of
holography. Liouville theory is central in two dimensional string theory.
It is believed to be dual, at least to all orders in string perturbation
theory, to a certain large $N$ matrix quantum mechanics (see \eg\ \gm\ for
a review). The two dimensional black hole appears in the near-horizon geometry
of $NS5$-branes and Calabi-Yau singularities \refs{\ov,\gk}
and thus plays a role in Little String Theory (see \eg\ \dk\ for a review).
An outstanding problem in two dimensional string theory and
LST is to understand microscopically the high energy density of states,
which was computed thermodynamically in \refs{\sahak,\kkk}. Studying
D-branes in these backgrounds might be useful for that.

\item{(4)} An important special case of (3) concerns the physics of
D-branes located near the tip of the two dimensional semi-infinite cigar
(or Euclidean two dimensional black hole). The near-horizon geometry
describing two $NS5$-branes intersecting along $\IR^{3,1}$ (or
equivalently a resolved conifold) is $\IR^{3,1}\times SL(2)/U(1)$.
D4-branes stretched between the fivebranes correspond in this geometry
to D-branes located near the tip of the cigar
\refs{\ElitzurPQ,\adg}. The low energy theory on
a collection of such branes is $N=1$ supersymmetric Yang-Mills theory \egk,
and one may hope that a better understanding of these D-branes will help
understand holography for $N=1$ SYM (see \malnun\ for a recent discussion).

\lref\stanone{S.~Stanciu, ``D-branes in an AdS(3) background,''
JHEP {\bf 9909}, 028 (1999) [hep-th/9901122].}
\lref\stantwo{J.~M.~Figueroa-O'Farrill and S.~Stanciu,
``D-branes in AdS(3) x S(3) x S(3) x S(1),''
JHEP {\bf 0004}, 005 (2000) [hep-th/0001199].}
\lref\BachasFR{
C.~Bachas and M.~Petropoulos, ``Anti-de-Sitter D-branes,''
JHEP {\bf 0102}, 025 (2001) [hep-th/0012234].}
\lref\FateevIK{V.~Fateev, A.~Zamolodchikov and A.~Zamolodchikov,
``Boundary Liouville field theory. I: Boundary state and boundary
two-point function,'' [hep-th/0001012].}
\lref\TeschnerMD{J.~Teschner, ``Remarks on Liouville theory with boundary,''
[hep-th/0009138].}
\lref\ZamolodchikovAH{A.~Zamolodchikov and A.~Zamolodchikov,
``Liouville field theory on a pseudosphere,'' [hep-th/0101152].}
\lref\tesch{J.~Teschner,
``On structure constants and fusion rules in the SL(2,C)/SU(2) WZNW  model,''
Nucl.\ Phys.\ B {\bf 546}, 390 (1999) [hep-th/9712256].}
\lref\tesss{J.~Teschner,
``Operator product expansion and factorization in the H-3+ WZNW model,''
Nucl.\ Phys.\ B {\bf 571}, 555 (2000) [hep-th/9906215].}
\lref\BehrendBN{
R.~E.~Behrend, P.~A.~Pearce, V.~B.~Petkova and J.~Zuber,
``Boundary conditions in rational conformal field theories,''
Nucl.\ Phys.\ B {\bf 570}, 525 (2000)
[Nucl.\ Phys.\ B {\bf 579}, 707 (2000)]
[hep-th/9908036].
}

\noindent
Unlike the case of D-branes in rational CFT's, which is well studied
(see \eg\ \BehrendBN\ for a review),
there has not been much work on non-compact interacting theories such
as $SL(2,R)$, with some notable exceptions. The authors of
\refs{\stanone,\stantwo} studied some aspects of the $SL(2,R)$ boundary
states; \ElitzurPQ\ discussed some properties of D-branes on the cigar.
The geometric, semi-classical interpretation of D-branes
in $AdS_3$ was clarified in \BachasFR. In particular, in this paper it was
shown that there are several classes of D-branes that preserve a diagonal
$SL(2)$ current algebra: pointlike D-instantons, Euclidean two dimensional
branes with worldvolume $H_2$, D-strings with worldvolume $AdS_2$ stretched
between two points on the boundary of $AdS_3$, and tachyonic D-strings with
worldvolume $dS_2$. We will reproduce these results below from an algebraic
analysis.

There was also some interesting work on D-branes in Liouville theory
\refs{\FateevIK,\TeschnerMD,\ZamolodchikovAH}, where it was shown that
there are two classes of branes in this case. The authors of
\refs{\FateevIK,\TeschnerMD} studied branes that are extended in the
Liouville direction, while \ZamolodchikovAH\ constructed localized branes.
The difference is reflected in the spectrum of open string excitations:
the extended branes of \refs{\FateevIK,\TeschnerMD} support open string
excitations which carry arbitrary momentum in the Liouville direction,
while the localized branes \ZamolodchikovAH\ have a finite number of
open string Virasoro primaries. It was also shown in
\refs{\FateevIK,\TeschnerMD,\ZamolodchikovAH} that the boundary of
the worldsheet is at a finite distance (in the dynamical worldsheet
metric) from the bulk for the extended branes, while for localized branes
this distance is infinite.

The above results on Liouville branes were obtained by solving the bootstrap
equations for the one point functions of bulk operators on the upper half
plane, and combining the resulting information with an analysis of the
annulus amplitude in the open and closed string channels
(``modular bootstrap'').
This seems to be a fruitful way of studying the theory.
The main purpose of this
paper is to repeat this analysis for the $AdS_3$ case. We will see that
the analogs of the localized branes of \ZamolodchikovAH\ in this case are
associated with D-instantons in $SL(2,R)$,
while the analogs of the extended branes of \refs{\FateevIK,\TeschnerMD} are
the $H_2$, $dS_2$ and $AdS_2$ branes of \BachasFR.
In the process we will learn more about these branes and answer some
of the questions mentioned above. In particular, we will see that
branes in $AdS_3$ introduce boundaries into the base space on which the
spacetime CFT is defined. Most of the detailed analysis will be done in
the Euclidean version of $AdS_3$, $H_3$; we will discuss the continuation
of the results to Lorentzian $AdS_3$.

The plan of the paper is as follows. In section 2 we start with a very brief
summary of $AdS_3$ CFT on the plane. We mainly establish the notation and
quote some results that are needed for the subsequent analysis. In section
3 we solve the bootstrap for the wavefunctions of localized D-branes satisfying
``symmetric gluing conditions'' $J^a(z)=\bar J^a(\bar z)$ for $z=\bar z$
($a=3,\pm$). We find an infinite number of solutions labeled by an integer
$r=1,2,3,\cdots$. The basic solution, $r=1$, does not contain any non-trivial
primary boundary operators. The spectrum of excitations contains the current
algebra block of the identity. This boundary state can be thought of as an
analog of the identity Cardy state for rational CFT's, or a pointlike
instanton in $AdS_3$. The spacetime theory in the background of a D-instanton
is equivalent to a two dimensional CFT on a manifold with a boundary.
The $r>1$ branes correspond to multi-instanton configurations, and
contain a finite number of degenerate $SL(2)$ primaries, whose properties
are very reminiscent of the $SU(2)$ case. This is analogous to the results
of \ZamolodchikovAH, who showed that the spectrum of excitations of the
localized branes in Liouville theory contains a finite number of degenerate
Virasoro representations, in a structure very reminiscent of minimal model
branes.

In section 4 we discuss the annulus amplitude corresponding to the
branes constructed in section 3, and show that the spectrum of boundary
operators proposed in section 3 leads to sensible modular properties.
This is used to determine the spectrum of open strings stretched between
different branes.

In section 5 we discuss a second class of solutions to the bootstrap
equations, describing two dimensional D-branes in $AdS_3$. We construct the
operator corresponding to the worldvolume electric field on the D-brane
and exhibit two classes of branes, one with a supercritical electric field,
and the other with subcritical field. In section 6 we discuss the corresponding
annulus amplitude and the spectrum of open strings living on these branes,
and strings connecting them to D-instantons.

In section 7 we discuss the generalization of the formalism to
``asymmetric'' gluing conditions for the $SL(2)$ currents, obtained
by twisting with automorphisms of the Lie algebra. We show that different
gluing conditions correspond in the spacetime CFT to boundaries at
different locations (related by conformal transformations).
In Minkowski space one can describe both spacelike and timelike
boundaries using different gluing conditions.

In section 8 we comment on the geometric interpretation of our results,
and in particular their relation to \BachasFR. We also discuss
possible extensions. Some technical results appear in the appendices.

\newsec{A brief review of conformal field theory on $AdS_3$}

Lorentzian $AdS_3$ can be thought of as a pseudosphere in $\IR^{2,2}$,
\eqn\pseudosp{x_1^2+x_2^2-x_0^2-x_3^2=-l^2,}
where $l$ is the radius of curvature of $AdS_3$.
A convenient parametrization of the space is via the coordinates
$(r,t,\theta)$,
\eqn\rttheta{\eqalign{
x_0=&\sqrt{l^2+r^2}\cos t\cr
x_3=&\sqrt{l^2+r^2}\sin t\cr
x_1=&r\cos\theta\cr
x_2=&r\sin\theta.\cr
}}
$r$ can be thought of as the radial coordinate on $AdS_3$, while
$(t,\theta)$ parametrize the boundary\foot{As usual, we will consider
the infinite cover of $SL(2,R)$, where $t$ is not periodic.}.
The metric on $AdS_3$ in these coordinates is
\eqn\metads{ds^2=\left(1+{r^2\over l^2}\right)^{-1}dr^2-
l^2\left(1+{r^2\over l^2}\right)dt^2+r^2d\theta^2.}
Another useful coordinate system is Poincare coordinates
$(u,\gamma,\bar\gamma)$, related to $(r,t,\theta)$ by the transformation
\eqn\primecoom{\eqalign{
u=&{1\over l}\left(\sqrt{l^2+r^2}\cos t + r\cos \theta\right) \cr
\gamma=&{\sqrt{l^2+r^2}\sin t+ r\sin \theta \over \sqrt{l^2+r^2}\cos t +
r\cos \theta} \cr
\bar \gamma=&{-\sqrt{l^2+r^2}\sin t+ r\sin \theta  \over \sqrt{l^2+r^2}
\cos t + r \cos \theta} .\cr}}
The metric \metads\ is
\eqn\metricupypm{ds^2=l^2({du^2\over u^2}+u^2d\gamma d\bar \gamma).}
The boundary of $AdS_3$ is at large $u$.
The coordinates \primecoom\ are obtained by parametrizing the
$SL(2,R)$ group manifold by the Gauss decomposition
\eqn\defg{g=\left(\matrix{
1&\bar\gamma \cr
0&1\cr}\right)
\left(\matrix{
u^{-1} &0\cr
0&u\cr}\right)\left(\matrix{
1&0\cr
\gamma &1\cr}\right)=
\left(\matrix{
\gamma \bar\gamma u +u^{-1} & \bar\gamma u\cr
\gamma  u & u \cr} \right).}
It is often convenient to analytically continue the geometry \metads\
to Euclidean space, \eg\ by setting $t=-i\tau$, or equivalently
replacing $x_3\to ix_3$ in \rttheta. This gives rise to a three
dimensional hyperbolic space $H_3\simeq SL(2,C)/SU(2)$. This is
the model that will be mostly studied below. We will parametrize $H_3$
as in \defg, with $\gamma$ a complex variable whose complex conjugate
is $\bar\gamma$. It should be kept in mind that the relation to
$SL(2)$ is through the analytic continuation that maps the coordinates
on $H_3$, $(u,\gamma,\bar\gamma)$, to \primecoom.

The WZNW model on $AdS_3$ is invariant under two copies of the
$SL(2,R)$ current algebra. The left moving symmetry is generated by
the currents $J^a(z)$, with $a=3,\pm$, satisfying the OPE algebra
\eqn\opealg{\eqalign{
J^3(z) J^\pm(w)\sim &{\pm J^\pm(w)\over z-w} \cr
J^3(z) J^3(w)\sim &-{{k\over2}\over (z-w)^2}\cr
J^-(z)J^+(w)\sim &{k\over (z-w)^2}+{2J^3(w)\over z-w}.\cr}}
A similar set of OPE's holds for the right moving $SL(2)$ current algebra.
The level $k$ of the current algebra \opealg\ is a real number, related to
$l$ in \metads\ via the relation $k=l^2$ (in string units). It
determines the central charge of the WZNW CFT via
\eqn\centch{c={3k\over k-2}.}
One is typically interested in $k>2$.

A natural set of observables is given by the eigenfunctions
of the Laplacian on $AdS_3$,
\eqn\limphin{\eqalign{
\Phi_h=&{1-2h\over\pi}\left({1 \over |\gamma-x|^2e^{Q\phi\over2}+
e^{-{Q\phi\over2}}}
\right)^{2h} = \cr
&-e^{Q(h-1)\phi} \delta^2(\gamma-x) + \CO(e^{Q(h-2)\phi}) +
{(1-2h)e^{-Qh\phi} \over \pi |\gamma-x|^{4h}} + \CO(e^{-Q(h+1) \phi}), }}
where we are using Poincare coordinates \metricupypm,
with\foot{Note that in \limphin\ we have rescaled $\phi$
and $\Phi_h$ relative to equations such as (2.8) in \ks.}
\eqn\defphi{u=e^{Q\phi\over2}.}
$Q$ is related to $k$ via
\eqn\Qk{Q^2={2\over k-2}\equiv-{2\over t}.}
The last equality defines
\eqn\defttt{t=-(k-2).}
$x$ is an auxiliary complex variable whose role
can be understood by expanding the operators $\Phi_h$ near the boundary
of $AdS_3$, $\phi\to\infty$, as is done on the second line
of \limphin. Note the difference
between the behavior for $h>1/2$ and $h<1/2$ \ks. For $h>1/2$,
the operators $\Phi_h$ are localized near the boundary at
$(\gamma,\bar\gamma)=(x,\bar x)$. For $h<1/2$, the delta function is
subleading, and the operators are smeared over the boundary. One can think of
$\Phi_h$ as the propagator of a particle with mass $h(h-1)$ from a point
$(x,\bar x)$ on the boundary, to a point
$(\phi,\gamma,\bar\gamma)$ in the bulk of $AdS_3$. Thus, $x$ labels the
position on the boundary of $AdS_3$, which is the base space of the CFT
dual to string theory on $AdS_3$ via the AdS/CFT correspondence.

\lref\ZamolodchikovBD{
A.~B.~Zamolodchikov and V.~A.~Fateev,
``Operator Algebra And Correlation Functions In The Two-Dimensional
Wess-Zumino SU(2) X SU(2) Chiral Model,''
Sov.\ J.\ Nucl.\ Phys.\  {\bf 43}, 657 (1986)
[Yad.\ Fiz.\  {\bf 43}, 1031 (1986)].}

The operators $\Phi_h$ are primary under the $\widehat{SL(2)}$
current algebra \opealg; they satisfy
\eqn\rrr{\eqalign{
J^3(z) \Phi_h(x,\bar x;w,\bar w)\sim &-{(x\partial_x+h)\Phi_h(x,\bar x)
\over z-w}\cr
J^+(z) \Phi_h(x,\bar x;w,\bar w)\sim &-{\left(x^2\partial_x+2hx\right)
\Phi_h(x,\bar x)\over z-w}\cr
J^-(z) \Phi_h(x,\bar x;w,\bar w)
\sim &-{\partial_x\Phi_h(x,\bar x)\over z-w}~.\cr
}}
Their worldsheet scaling dimensions are
\eqn\scdim{\Delta_h=-{h(h-1)\over k-2}={h(h-1)\over t}~.}
It is very convenient \ZamolodchikovBD\ to ``Fourier transform'' the
$SL(2)$ currents as well, and define
\eqn\intoppp{J( x; z)\equiv -J^+(x; z)=2 x J^3(z)- J^+(z) - x^2 J^-(z)~.}
Since $J_0^-=-\partial_x$ is the generator of translations in $x$
(see \rrr) we can think of \intoppp\ as a result of ``evolving'' the
currents $J^a(z)$ in $x$:
\eqn\jofxdef{\eqalign{ J^+(x;z)=&e^{-xJ_0^-} J^+(z) e^{xJ_0^-}=J^+(z)
-2xJ^3(z)+x^2J^-(z)\cr
J^3(x;z)=&e^{-xJ_0^-} J^3(z) e^{xJ_0^-}=J^3(z) -xJ^-(z) =
-\half \partial_x J^+(x;z)\cr
J^-(x;z)=&e^{-xJ_0^-} J^-(z) e^{xJ_0^-}=J^-(z)
=\half \partial_x^2 J^+(x;z).\cr}}
The OPE algebras \opealg\ and \rrr\ can be written in terms of $J(x;z)$
as follows:
\eqn\JJ{J( x; z) J( y; w)\sim k {(y-x)^2\over( z- w)^2}+ { 1\over
z- w}\left[(y-x)^2\partial_y -2(y-x) \right] J( y; w)}
\eqn\JPhi{J( x; z) \Phi_h(y,\bar y;w,\bar w)\sim  {1 \over z-w} \left[(
y-x)^2\partial_y +2h(y-x)\right]\Phi_h(y, \bar y) .}

\noindent
It is sometimes useful to expand the operators \limphin\ in modes,
\eqn\ppp{\Phi_h(x,\bar x)=\sum_{m,\bar m}V_{h-1;m,\bar m} x^{-m-h}
\bar x^{-\bar m-h}}
or
\eqn\vjmbarm{V_{j;m,\bar m}=\int d^2x x^{j+m}\bar x^{j+\bar m}
\Phi_{j+1}(x,\bar x)~.}
Note that \rrr\ implies that $V_{j;m,\bar m}$ transforms under
$\widehat{SL(2)}$ as follows:
\eqn\vjmpm{\eqalign{
J^3(z)V_{j;m,\bar m}(w)=&{m\over z-w}
V_{j;m,\bar m}\cr
J^{\pm}(z)V_{j;m,\bar m}(w)=&{(m\mp j)\over z-w}
V_{j;m\pm 1,\bar m}.\cr}}

\noindent
As is clear from \scdim, the operators $\Phi_h$ and $\Phi_{1-h}$ are
closely related. They are related by a reflection symmetry \tesch,
\eqn\teschone{\Phi_h(x,\bar x;z,\bar z)=\RR(h){2h-1\over\pi}\int d^2x'
|x-x'|^{-4h}\Phi_{1-h}(x',\bar x';z,\bar z)}
where the $x'$ integral runs over the complex plane.
The reflection coefficient $\RR(h)$ depends on the normalization
of the operators (see \gksch). In the normalization used in \gksch,
which we will adopt here, it is equal to
\eqn\rrhh{\RR(h)={\Gamma(1+{2h-1\over t})\over \Gamma(1-{2h-1\over t})}~.}
Note that in the semiclassical limit $t\to-\infty$ the reflection coefficient
goes to one; the resulting semiclassical relation \teschone\ can be verified
directly by using \limphin.

\newsec{$SL(2)$ conformal field theory on the upper half plane (I)}

After reviewing $AdS_3$ CFT on the plane in the previous section,
we turn next to the construction of D-branes in this background.
We would like to analyze the theory on the upper half plane
${\rm Im} z\ge 0$, with boundary conditions that preserve conformal
symmetry. We will impose the standard requirement on the
worldsheet stress tensor,
\eqn\stressbound{T(z)=\bar T(\bar z);\qquad{\rm for}\; z=\bar z.}
We will furthermore require that the D-branes preserve a diagonal
$SL(2)$ current algebra. The latter is not necessary -- one
certainly expects to find D-branes that do not satisfy this
requirement. Nevertheless, it is useful to analyze the most
symmetric D-branes before moving on to less symmetric ones.

The simplest boundary conditions on the currents are:
\eqn\symcur{J^a(z)=\bar J^a(\bar z);\qquad{\rm for}\; z=\bar z,
\qquad a=3,+,-~.}
In this section we will analyze localized D-branes that
arise when we impose \stressbound, \symcur. We will follow
closely the discussion of \refs{\ZamolodchikovAH}.

\lref\CardyIR{J.~L.~Cardy,
``Boundary Conditions, Fusion Rules And The Verlinde Formula,''
Nucl.\ Phys.\ B {\bf 324}, 581 (1989).}

Consider the one point function of the bulk observable \limphin\
on the upper half plane,
\eqn\llaa{\langle\Phi_h(x,\bar x;z,\bar z)\rangle
={U(h)\over(x-\bar x)^{2h}(z-\bar z)^{2\Delta_h}}~.}
The $z$ dependence in \llaa\ is fixed by the conformal
symmetry which remains unbroken on the upper half plane
\stressbound. The $x$ dependence follows from the unbroken
$SL(2)$ symmetry\foot{Since $J^a+\bar J^a$ is a conserved
charge, we see from \rrr\ that the correlator \llaa\ must satisfy
$\partial_x+\partial_{\bar x}=0$, $x\partial_x+\bar x\partial_{\bar x}+2h=0$,
$x^2\partial_x+\bar x^2\partial_{\bar x}+2h(x+\bar x)=0$, whose solution
is const$\times (x-\bar x)^{-2h}$.} \symcur. As is standard in boundary
conformal field theory (BCFT), one can think of the one point functions
\llaa\ as the wavefunctions of the corresponding boundary states \CardyIR.
We would like to determine them, and use that to characterize the boundary
states.

Note that the one point function \llaa\ exhibits a singularity
as $x\to\bar x$. We will see below that the meaning of this is the following.
Recall that $x$ labels the base space on which the CFT dual to string theory
on $AdS_3$ lives \refs{\gks,\ks}. Introducing a boundary with the boundary conditions
\stressbound, \symcur\ into the worldsheet corresponds in the spacetime CFT
to the appearance of
a boundary at $x=\bar x$ with conformal boundary conditions similar to
\stressbound\ for the spacetime stress tensor constructed in \refs{\gks,\ks}.
The singularity as $x\to\bar x$ in \llaa\ corresponds to the operator
$\Phi_h$ approaching the boundary {\it in spacetime}.

It is also useful to note that $U(h)$ has a simple transformation under
$h\to 1-h$. To derive it we use the reflection symmetry \teschone.
Take the expectation value of both sides of \teschone. This gives
\eqn\testhree{{U(h)\over (x-\bar
x)^{2h}}=\RR(h){2h-1\over\pi}U(1-h) \int d^2x'{|x-x'|^{-4h}\over (x'-\bar
x')^{2(1-h)}}}
The $x'$ integral runs over the complex plane. Denote
\eqn\notxxprime{\eqalign{
x=&x_1+ix_2\cr
x'=&x_1'+ix_2'\cr
}}
with $x_2>0$, and $x_2'$ running from $-\infty$ and $\infty$. By shifting
$x_1'$ we can get rid of the $x_1$ dependence on the r.h.s. of \testhree,
which is consistent with the fact that the l.h.s. is independent of $x_1$.
We then find
\eqn\dointeg{\eqalign{
\int d^2x'{|x-x'|^{-4h}\over (x'-\bar x')^{2(1-h)}}=&
\int_{-\infty}^\infty dx_1'\int_0^\infty dx_2'
{\left[(x_1')^2+(x_2'-x_2)^2\right]^{-2h}
\over (2e^{i\pi\over2}x_2')^{2(1-h)}}+\cr
+&\int_{-\infty}^\infty dx_1'\int_{-\infty}^0 dx_2'
{\left[(x_1')^2+(x_2'-x_2)^2\right]^{-2h}
\over (2e^{i\pi\over2}x_2')^{2(1-h)}}\cr
}}
The $x_1'$ integral is performed by rescaling $x_1'=z|x_2'-x_2|$.
One gets (apart from a power of $|x_2'-x_2|$)
\eqn\aahhaa{2A_h\equiv2\int_0^\infty dz(1+z^2)^{-2h}={\Gamma(2h-\half)
\Gamma(\half)\over\Gamma(2h)}}
The remaining $x_2'$ integral looks like
\eqn\upphalf{
\int_0^\infty dx_2'{|x_2'-x_2|^{1-4h}\over(2e^{i\pi\over2}x_2')^{2(1-h)}}}
for the upper half plane, and
\eqn\lowhalf{
\int_{-\infty}^0
dx_2'{|x_2'-x_2|^{1-4h}\over(2e^{i\pi\over2}x_2')^{2(1-h)}}}
for the lower half plane. The contribution of the upper half plane
vanishes. It is equal to
\eqn\contrupp{{2A_h\over x_2^{2h}(2e^{i\pi\over2})^{2(1-h)}}
\int_0^\infty dx_2'(x_2')^{2(h-1)}|1-x_2'|^{1-4h}}
and by using the standard Euler integral
\eqn\venform{\int_0^1 dx x^{a-1}(1-x)^{b-1}={\Gamma(a)\Gamma(b)\over
\Gamma(a+b)},}
one finds that this vanishes.
The contribution from the lower half plane is non-zero; it is given by
\eqn\contrlow{{2A_h\over x_2^{2h}(2e^{-{i\pi\over2}})^{2(1-h)}}
\int_{-\infty}^0 dx_2'(-x_2')^{2(h-1)}(1-x_2')^{1-4h}.}
By using \venform\ again, and also
\eqn\gammaiden{\eqalign{
&\Gamma(\half)=\sqrt{\pi}\cr
&\Gamma(2x)={1\over\sqrt{\pi}}2^{2x-1}\Gamma(x)\Gamma(x+\half),\cr
}}
one finds that
\eqn\uuuhhh{U(h)=-\RR(h)U(1-h)=-
{\Gamma(1+{2h-1\over t})\over \Gamma(1-{2h-1\over t})}U(1-h).}
Defining\foot{The $(-)^{\Delta_h}$ is irrelevant for the
present discussion since $\Delta_h=\Delta_{1-h}$, but it will
be convenient later.}
\eqn\fhdef{f(h)={(-)^{\Delta_h}U(h)\over \Gamma(1+{2h-1\over t})}}
we have
\eqn\fhsymm{f(h)=-f(1-h).}

\noindent
To determine $f(h)$ we apply a procedure that was used for
Liouville theory in \ZamolodchikovAH. The $SL(2)$ CFT has
degenerate operators of the form \limphin\
with (see \tesch\ for more details)
\eqn\degops{h_{r,s}={1-r\over2}-{1-s\over2}t}
where $r,s=1,2,3,\cdots$. For irrational $k$, the Fock module corresponding
to $h_{r,s}$ contains a single null state at level $r(s-1)$. Consider, for
example, the special case $s=1$. The degenerate representations have
$h_{r,1}=(1-r)/2$, and the null state is at level zero.
Looking back at \limphin\ this is natural: $\Phi_{h_{r,1}}$
is in this case a polynomial of degree $r-1$ in $x$ and $\bar x$,
and the null state is
\eqn\nullrsz{\partial_x^r\Phi_{(1-r)/2}=\partial_{\bar x}^r\Phi_{(1-r)/2}=0,
\qquad r=1,2,3,\cdots.}
The operators $\Phi_{(1-r)/2}$ correspond to finite $r$ dimensional
representations of $SL(2)$. They are direct generalizations of the
finite dimensional spin $(r-1)/2$ representations of $\widehat{SU(2)}$
which are studied in the language used here in \ZamolodchikovBD.

Equation \nullrsz\ gives strong constraints on the OPE of the degenerate
operators with generic operators $\Phi_h$. For example, as reviewed
in \gksch, the first non-trivial operator $\Phi_{-\half}$ satisfies
\eqn\opehalf{\Phi_{-\half}(x)\Phi_h(y)=
C_-(h)\Phi_{h-\half}(y)+|x-y|^2C_+(h)\Phi_{h+\half}(y)+\cdots}
where the ``$\cdots$'' stand for current algebra descendants, and
we have suppressed the dependence on worldsheet locations of the operators,
which can be easily restored using \scdim. The structure constants
$C_\pm$ are given by (in our normalizations; see \gksch\ for a more detailed
discussion)
\eqn\cplusminus{
\eqalign{
C_+(h)=&{2\over\pi}\RR(-\half)\cr
C_-(h)=&{2\over\pi}\RR(-\half)
{\Gamma(-{2(h-1)\over t})\Gamma(1+{2h-1\over t})\over
\Gamma(1+{2(h-1)\over t})\Gamma(-{2h-1\over t})}.\cr}}
In what follows, we would like to use the degenerate operators
\degops\ to obtain constraints on the one point function of operators
with generic $h$, \llaa.

To achieve that, consider the two point function on the upper half plane:
\eqn\lla{G_{-{1\over2}}(h)=
\langle\Phi_{-\half}(x,\bar x;z,\bar z)\Phi_h(y,\bar y;w,\bar w)\rangle}
Using the unbroken worldsheet conformal symmetry \stressbound, and
$SL(2)$ symmetry \symcur, one can write the two point function as follows:
\eqn\llb{G_{-{1\over2}}(h)={(w-\bar w)^{-2\Delta_h}\over
(z-\bar z)^{2\Delta_{-\half}}}{x-\bar x\over(y-\bar y)^{2h}}
\CF(\eta_{\rm ws},\eta_{\rm st})}
where $\eta_{\rm ws}$ and $\eta_{\rm st}$ are the cross ratios
on the worldsheet and in spacetime,
\eqn\llc{\eqalign{
\eta_{\rm ws}=&{|z-w|^2\over(z-\bar z)(w-\bar w)}\cr
\eta_{\rm st}=&{|x-y|^2\over(x-\bar x)(y-\bar y)}.\cr}}
Consider the dependence of $G_{-\half}$ on $x$.
Since the operator $\Phi_{-\half}$ satisfies the differential
equation \nullrsz\ with $r=2$, the right hand side of \llb\
must be a linear function of $x$ and of $\bar x$.
This means that the function $\CF$ is in fact linear in $\eta_{\rm st}$:
\eqn\lld{\CF(\eta_{\rm ws}, \eta_{\rm st})=a_0(\eta_{\rm ws})+
\eta_{\rm st}a_1(\eta_{\rm ws})~.}
As is well known, the two point function of bulk operators on the
upper half plane is closely related to a four point function (of the
operators and their mirror images) on the plane. In particular, the
function of the cross ratio $\CF$ \llb, \lld\ is given by a combination
of the current algebra blocks on the sphere. Due to \opehalf\ there are
in this case two blocks, and one has
\eqn\tfre{\CF(\eta_{\rm ws}, \eta_{\rm st})=
a_-(\CF_0^{(-)}+\eta_{\rm st}\CF_1^{(-)})+
a_+(\CF_0^{(+)}+\eta_{\rm st}\CF_1^{(+)}),}
where $\CF_{0,1}^{(\pm)}$ are given in \gksch. We present them
here for completeness:
\eqn\solone{\eqalign{
&\CF_0^{(-)}
=x^{-a}(1-x)^{-a}F(-2a,2b-2a; b-2a;x)\cr
&=x^{-a}(1-x)^{a-b}F(b,-b;b-2a;x)\cr
&\CF_1^{(-)}={2a\over b-2a}x^{-a}(1-x)^{-a}F(1-2a,2b-2a;b-2a+1;x)\cr
&={2a\over b-2a}x^{-a}(1-x)^{a-b}F(b,1-b;b-2a+1;x)\cr
}}
\eqn\soltwo{\eqalign{
&\CF_0^{(+)}=x^{a-b+1}(1-x)^{a-b}F(2a-2b+1,2a+1; 2a-b+2;x)\cr
&=x^{a-b+1}(1-x)^{-a}F(1-b,1+b; 2a-b+2;x)\cr
&\CF_1^{(+)}={b-2a-1\over b}x^{a-b}(1-x)^{-a}
F(1-b,b;2a-b+1;x)\cr
&={b-2a-1\over b}x^{a-b}(1-x)^{a-b}F(2a-2b+1,2a;2a-b+1;x),\cr
}}
where we used the notation
\eqn\defsol{a\equiv {h\over t};\;\;b\equiv{1\over t};\;\;
x\equiv\eta_{\rm ws},}
and $F$ is the hypergeometric function (see appendix A for some of its
properties).

To determine the constants $a_\pm$ in \tfre, consider the two point
function \lla\ in the limit $z\to w$ (\ie\ $\eta_{\rm ws}\to 0$).
It is not difficult to see that $\CF^{(-)}$ corresponds to the
contribution of the block of $\Phi_{h-\half}$, while $\CF^{(+)}$
is the contribution of $\Phi_{h+\half}$ and its descendants.
Taking the limit $\eta_{\rm ws}\to 0$ in \solone, \soltwo\ and using the
relations
\eqn\dddab{\eqalign{\Delta_{h-\half}-\Delta_{-\half}-\Delta_h=&
-{h\over t}=-a\cr
\Delta_{h+\half}-\Delta_{-\half}-\Delta_h=&
{h-1\over t}=a-b\cr}}
one furthermore
finds that the contribution of the primary $\Phi_{h-\half}$ comes from the
leading term in $\CF_0^{(-)}$, while that of $\Phi_{h+\half}$ comes from
$\CF_1^{(+)}$. The leading terms in $\CF_1^{(-)}$ and $\CF_0^{(+)}$ are
the contributions of $\partial_y\Phi_{h-\half}(y;w)$ and
$J\Phi_{h+\half}(y;w)$, respectively.

Using the OPE \opehalf\ and the behavior of $\CF^{(\pm)}$ at small
$\eta_{\rm ws}$, we conclude that
\eqn\aplusminus{\eqalign{
a_-=&C_-(h)U(h-\half)\cr
a_+=&{1\over 1-2h-t}C_+(h)U(h+\half).\cr
}}
Plugging this into \tfre\ and comparing to \lld\ we see that
\eqn\lle{
\eqalign{
a_0(\eta_{\rm ws})=&C_-(h)U(h-\half)\CF_0^{(-)}+
{1\over 1-2h-t}C_+(h)U(h+\half)\CF_0^{(+)}\cr
a_1(\eta_{\rm ws})=&C_-(h)U(h-\half)\CF_1^{(-)}+
{1\over 1-2h-t}C_+(h)U(h+\half)\CF_1^{(+)}\cr}}
where $\CF^{(\pm)}$ are the blocks of \solone, \soltwo.
The resulting equation for the two point function $G_{-{1\over2}}(h)$
involves known quantities such as the structure constants \cplusminus\ and
the conformal blocks \solone, \soltwo, and the unknown one point function
$U(h)$, \llaa.

A non-trivial constraint on $U(h)$ comes from considering the two point
function \lla\ in the limit $z\to\bar z$, $w\to\bar w$.
In this limit the bulk operators $\Phi_{-\half}$ and $\Phi_h$ approach
the boundary of the worldsheet and we expect them to create boundary
operators $\Psi_{\tilde h}$. It is natural to expect that these
operators are restricted to the boundary {\it in spacetime} as well.
This is an important general feature. In addition to the bulk observables
\limphin, in the presence of D-branes one finds operators $\Psi_h(x;z)$
which live on the boundary of the worldsheet ${\rm Im}z=0$, and of spacetime
${\rm Im}x=0$. We will explain this further below. For now we
note that the worldsheet boundary scaling dimension of $\Psi_h$ is
\eqn\bounddim{\Delta_h^{(b)}={h(h-1)\over t}}
while the spacetime boundary scaling dimension is $h$.

Returning to the two point function \lla, clearly in the limit
$z\to\bar z$ there should still be only two blocks.
To describe them explicitly,
consider the operator $\Phi_{-\half}$. As mentioned above, the mode
expansion \ppp\ truncates:
\eqn\modehalf{\Phi_{-\half}(x,\bar x;z,\bar z)=\sum_{m,\bar m=-\half,\half}
V_{-\frac32;m,\bar m}(z,\bar z)x^{\half-m}\bar x^{\half-\bar m}.}
Near the boundary of the worldsheet, $\Phi_{-\half}$ \modehalf\ should be
expanded in boundary operators, both on the worldsheet and in spacetime.
The spacetime expansion is particularly simple; one can write
\eqn\expphihalf{\Phi_{-\half}(x,\bar x)=(x-\bar x)V_1+[V_2+(x+\bar x)V_3+
x\bar xV_4]}
where $V_1,\cdots V_4$ are combinations of $V_{-\frac32;m,\bar m}$ evaluated
at $z\to\bar z$. In the limit $x\to\bar x$ one can think of \expphihalf\ as
follows:
\eqn\llg{\Phi_{-\half}(x\simeq\bar x)\sim
A_0(x-\bar x)+A_1\Psi_{-1}(x;z)}
The first term on the r.h.s. corresponds to the identity operator
on the boundary, with $A_0$ the corresponding structure constant
and $(x-\bar x)$ taking care of the spacetime scaling dimension.
In the second term, $\Psi_{-1}$ is a quadratic polynomial in $x$,
corresponding to the finite dimensional spin one representation of
$SL(2)$, and $A_1$ is the relevant structure constant. $\Psi_{-1}$
satisfies the boundary analog of \nullrsz. It belongs to an infinite
set of boundary operators degenerate at level zero, $\Psi_{(1-r)/2}$,
which satisfy
\eqn\nullrszbo{\partial_x^r\Psi_{(1-r)/2}=0,\qquad r=1,2,3,\cdots.}
Note that group theoretically, \llg\ is simply the statement that
multiplying two spin $\half$ $SL(2)$ representations (corresponding
to $\Phi_{-\half}$ and its mirror image) gives representations with
spins zero and one, $\Psi_0=1$ and $\Psi_{-1}$.

To reiterate, when $\Phi_{-\half}$ approaches the boundary of the worldsheet
in \lla, it can be expanded in boundary operators (both on the worldsheet
and in spacetime). There are two terms in the expansion, corresponding to
the degenerate operators $1$ and $\Psi_{-1}$. Similarly, when $\Phi_h$
approaches the boundary, it can be expanded in boundary operators; the two
point function \lla\ is only sensitive to the contributions of the boundary
operators $1$ and $\Psi_{-1}$ in this expansion.

We will next obtain a constraint on the one point function $U(h)$ \llaa\
by computing the contribution of the identity operator to the two point
function \lla\ in two different ways. One is to use the explicit form
\llb\ -- \lle\ in the limit $z-\bar z\to i0^+$, $w-\bar w\to i0^+$, \ie\
$\eta_{\rm ws}\to-\infty$. The contribution of the identity operator
in this limit is given by the constant term in
$\CF(\eta_{\rm ws},\eta_{\rm st})$. Using equations \llb\ -- \lle\ and
the properties of hypergeometric functions reviewed in appendix A, we find
\eqn\finf{\eqalign{\CF(\eta_{\rm ws}\to -\infty)=&
(-1)^{-a}C_-(h)U(h-\half)
{\Gamma(b-2a)\Gamma(2b)\over \Gamma(b)\Gamma(2b-2a)}\cr
&+(-1)^{a-b}C_+(h)U(h+\half)
{\Gamma(2a-b+1)\Gamma(2b)\over \Gamma(b)\Gamma(2a+1)}
+\cdots\cr}}
where ``$\cdots$'' stand for $\eta$ dependent terms, and $a,b$
are given in \defsol.
On the other hand, the contribution of the identity operator to
\lla\ factorizes as a product of one point functions,
\eqn\ppuu{\eqalign{\langle
\Phi_{-\half}(x,\bar x;z,\bar z)\Phi_h(y,\bar y;w,\bar w)\rangle
\simeq & \langle \Phi_{-\half}(x,\bar x;z,\bar z)\rangle
\langle \Phi_h(y,\bar y;w,\bar w)\rangle\cr
= & {(w-\bar w)^{-2\Delta_h}\over
(z-\bar z)^{2\Delta_{-\half}}}{x-\bar x\over(y-\bar y)^{2h}}
U(-\half)U(h)\cr}}
where in the last equality we used \llaa\ twice.
Some comments are in order here:
\item{(1)} The factorization property \ppuu\ is the $SL(2)$ analog
of eq. (2.12) in \ZamolodchikovAH\ for the Liouville case.
It should be emphasized that it is important for its validity
that the spectrum of states that live on the boundary is discrete
(and as we will see, it is even finite). We will later encounter branes
for which the open string spectrum is continuous, and there the
relation \ppuu\ will not hold.
\item{(2)} In the derivation of \ppuu, the one point function $U(h)$
is defined by \llaa, where the correlator is normalized by dividing
by the partition sum. Naively this means that $U(0)=1/\pi$ (since
$\Phi_0=1/\pi$; see \limphin).
However, as discussed above and in \gksch, we are using a normalization
in which operators with $h<1/2$ are renormalized by a factor of $\RR(h)$
\rrhh\ compared to their semiclassical form. Thus, we in fact expect
\eqn\uhnorm{U(0)={\langle\Phi_0\rangle\over\langle 1\rangle}={1\over\pi}
\RR(0)={1\over\pi}{\Gamma(1-{1\over t})\over\Gamma(1+{1\over t})}.}

\noindent
We now have two different expressions for the contribution of the identity
to the two point function \lla\ in the limit $\eta_{\rm ws}\to-\infty$.
One is obtained by plugging \finf\ in \llb; the other is \ppuu. Equating
the two, using the value of $C_{\pm}$ found in \cplusminus, and writing
the relation in terms of $f(h)$ \fhdef, we find
\eqn\fsimple{\pi\Gamma(1+{1\over t})f(-\half)f(h)=
f(h-\half)+f(h+\half),}
where $f(h)=-f(1-h)$ \fhsymm.

Equation \fsimple\ has many solutions.
We will next show that when one includes
a similar constraint on the one point function coming from the degenerate
operator $\Phi_{t\over 2}$ (corresponding to $r=1,s=2$ in \degops), $f(h)$
is fixed uniquely.

The degenerate operator $\Phi_{t\over2}$ is discussed in
\refs{\tesch,\gksch}. As shown in these papers, the combination
\eqn\abc{\theta(x)=\half t(t+1)\partial_x^2 J \Phi_{t\over2}+
(t+1)\partial_xJ\partial_x\Phi_{t\over2}+J\partial_x^2\Phi_{t\over2},}
which is manifestly a current algebra descendant of $\Phi_{t\over2}$,
is also primary. In a unitary theory, this would imply that $\theta$
should be set to zero. $SL(2)$ CFT is not unitary, but it is believed
that one should still set $\theta=0$. In some sense, this is part of
the definition of the theory \tesch.

The vanishing of $\theta$ imposes a constraint on the OPE's of
$\Phi_{t\over2}$ with other observables. As described in
\refs{\tesch,\gksch}, one has
\eqn\ooe{\Phi_{t\over2}(x)\Phi_h(y)=C_1(h)\Phi_{h+{t\over2}}(y)+
{C_2(h)\over|x-y|^{2t}}\Phi_{h-{t\over2}}(y)+{C_3(h)\over|x-y|^{2(t+2h-1)}}
\Phi_{1-{t\over2}-h}}
where the structure constants are:
\eqn\strconst{\eqalign{
C_1(h)=&-{t^2(1-t)\over \pi(2h+t-1)^2}\RR({t\over2})\cr
C_2(h)=&{1-t\over\pi}\RR({t\over2})\cr
C_3(h)=&{1\over\pi}(2h-1)(1-t)\RR({t\over2})
{\Gamma(1+{2h-1\over t})\Gamma(1-2h)\Gamma(1-t)\Gamma(2h+t-1)
\over\Gamma(1-{2h-1\over t})\Gamma(2h)\Gamma(t)\Gamma(2-2h-t)}.\cr}}

\noindent
To derive a constraint on the one point function $U(h)$, we proceed
in the same way as before. Consider the two point function on the upper
half plane,
\eqn\llttwo{G_{t\over2}(h)=
\langle\Phi_{t\over2}(x,\bar x;z,\bar z)\Phi_h(y,\bar y;w,\bar w)\rangle.}
The dependence on the worldsheet and spacetime positions is
\eqn\llbt{G_{t\over2}(h)={(w-\bar w)^{-2\Delta_h}\over
(z-\bar z)^{2\Delta_{t\over2}}}{(x-\bar x)^{-t}\over(y-\bar y)^{2h}}
\CF(\eta_{\rm st},\eta_{\rm ws})}
where $\eta_{\rm ws}$, $\eta_{\rm st}$ are given by \llc.
Since the OPE \ooe\ contains three $SL(2)$ representations,
the function $\CF$ is given in this case by
\eqn\diskconfbl{\CF(x;z)=a_AF_A+a_BF_B+a_CF_C,}
where $F_{A,B,C}$ are the current algebra blocks corresponding
to $\Phi_{h+{t\over2}}$, $\Phi_{h-{t\over2}}$ and $\Phi_{1-h-{t\over2}}$,
respectively. They are given by (see \refs{\tesch,\gksch})
\eqn\fabc{\eqalign{
F_A(x;z)=&z^h(1-z)^hF_1(2h,t,2h+t-1;2h+t;x,z)\cr
F_B(x;z)=&x^{-t}z^{1-h}(1-z)^hF_1(t,t,1-t;2-2h;{z\over x},z)\cr
F_C(x;z)=&z^h(1-z)^he^{-i\pi(1-t)}{\Gamma^2(2h)\over\Gamma(2h+1-t)
\Gamma(2h+t-1)}\times\cr
&\left[Z_8-{\Gamma(2h+1-t)\Gamma(1-2h-t)
\over\Gamma^2(1-t)}e^{2i\pi h}Z_1\right]\cr
}}
where
\eqn\zeightone{\eqalign{
Z_8=&x^{-2h}F_1(2h,1-t,t+2h-1;2h+1-t;{1\over x}, {z\over x})\cr
Z_1=&F_1(2h,t,t+2h-1;t+2h;x,z).\cr
}}
In the last equations, $F_1$ is a hypergeometric function
in two variables defined in appendix A. The structure constants
$a_A$, $a_B$, $a_C$ are determined as before by studying \llttwo\
in the limit $z\to w$, $x\to y$ \ie\ $\eta_{\rm ws},\eta_{\rm st}\to 0$.
The small $x,z$ behavior of $\CF$ leads to
\eqn\aabbccdd{\eqalign{
a_A=&C_1(h)U(h+{t\over2})\cr
a_B=&C_2(h)U(h-{t\over2})\cr
a_C=&C_3(h)U(1-h-{t\over2}).\cr
}}
If we now send the cross ratios to $-\infty$, we
find a non-linear equation for $U(h)$ \llaa, as before.

In order to do that we need the behavior of the functions
$F_A$, $F_B$, $F_C$ at $z\to-\infty$, $x\to-\infty$
(in this order of limits). Moreover, we are only interested in
the constant term which gives the identity block in the t-channel.
This term can be computed by using certain relations satisfied
by hypergeometric functions in two variables, described in
\ref\AAKK{P. Appell and J. Kampe de Feriet, ``Fonctions
Hypergeometriques et Hyperspheriques-Polynomes d'Hermite,''
Gauthier-Villars, Paris, 1926.}. We outline the calculation in
appendix B. We find
\eqn\behinf{\eqalign{
F_A(x\to-\infty;z\to-\infty)\simeq &(-)^h{2h+t-1\over t-1}\cr
F_B(x\to-\infty;z\to-\infty)\simeq &(-)^{-h}{2h-1\over t-1}\cr
F_C(x\to-\infty;z\to-\infty)\simeq &(-)^h{\Gamma(2h)\Gamma(2-2h-t)\Gamma(t-1)
\over\Gamma(1-2h)\Gamma(2h+t-1)\Gamma(1-t)}.\cr
}}
Using \behinf\ in \diskconfbl\ and equating the result to the
analog of \ppuu\ for this case, we get the relation:
\eqn\fhttwosim{\pi\Gamma(1+{1\over t})f({t\over2})f(h)
=f(h+{t\over2})+f(h-{t\over2})+f(1-h-{t\over2}).}
The reflection symmetry \fhsymm\ allows one to simplify
\fhttwosim\ considerably:
\eqn\fhttfinal{\pi\Gamma(1+{1\over t})f({t\over2})f(h)
=f(h-{t\over2})}
This relation\foot{which, together with \fhsymm\ implies that
$f(h+t)=f(h)$.} together with \fsimple\ determines the function $f$ uniquely,
at least for irrational $t$ (and then by analytic continuation).
The solution is
\eqn\solnfh{f(h)=\rho(t){\sin\left[{\pi\over t}(2h-1)(2h'-1)\right]\over
\sin\left[{\pi\over t}(2h'-1)\right]}}
where
\eqn\rhottt{\rho(t)=-{1\over\pi\Gamma(1+{1\over t})}}
and~\foot{The denominator in \solnfh\ has singularities for rational $t$.
These singularities are associated with the vanishing of the disk partition
sum for the corresponding boundary states labeled by $h'$. One possible
interpretation of this is that for rational $t$ one should restrict to a finite
subset of the branes constructed here; \eg\ for $t\in Z$ one would only
keep branes with $|2h'-1|<|t|$. It would be interesting to investigate
this further.}
\eqn\intsol{0\neq 2h'-1\in Z~.}
Note that this solution is normalized such that
\eqn\uhone{\eqalign{
U(h=1)=&-{1\over\pi}\cr
U(h=0)=&{1\over\pi}{\Gamma(1-{1\over t})\over\Gamma(1+{1\over t})}\cr
}}
in agreement with the expectation \uhnorm.

To summarize, the conclusion of the analysis is that boundary states are
labeled by a positive integer $|2h'-1|$ and have wavefunctions given by
\solnfh. These boundary states are in one to one correspondence with
degenerate operators of the form $\Phi_{-j'}$ \nullrsz, with
$-h'=j'=0,\half,1,\cdots$.

Now that we understand the one point functions $U(h)$, we
can compute the amplitude for creating $\Psi_{-1}$ on the boundary
(see \llg). In analogy to
\ZamolodchikovAH\ we expect the following to happen:
for $|2h'-1|>1$ we should get a non-zero amplitude for this, while
for $|2h'-1|=1$ the amplitude should vanish. More generally,
we expect that for $h'=-j$ the fundamental strings
that connect the brane to itself contain only the degenerate
operators $\Psi_{-j'}$, with
$j'$ running from zero up to $2j$ in jumps of one (and of course
their current algebra descendants).

To see that this general picture is indeed consistent with our
results, consider again the two point function \lla. As we see
from \llg, to study the appearance (or lack thereof) of $\Psi_{-1}$
in the limit $x\to\bar x$, we should look for a constant term
(in $x-\bar x$) in $G_{-\half}(h)$. This constant term will then
measure the product of:
(1) the amplitude to create $\Psi_{-1}$ from $\Phi_{-\half}\to$
boundary; (2) the amplitude to create $\Psi_{-1}$ from $\Phi_h\to$
boundary; (3) the boundary two point function
$\langle\Psi_{-1}\Psi_{-1}\rangle$.

As is clear from the form of the two point function \llb, we are
looking for a term that goes like $\eta_{\rm st}$ in $\CF$, \ie\
we are interested in $a_1$ (see \lld, \lle). Thus, we are interested
in the behavior of $\CF_1^{(\pm)}$ as $\eta_{\rm ws}\to-\infty$.
By using the explicit forms \solone, \soltwo, we find\foot{After
projecting out the contribution of the identity sector.}
\eqn\largeetaf{\eqalign{
\CF^{(-)}_1\simeq&(-)^{1+{h\over t}}(-\eta_{\rm ws})^{-{2\over t}}
{\Gamma({1-2h\over t})\Gamma(1-{2\over t})\over
\Gamma(1-{1\over t})\Gamma(-{2h\over t})}\cr
\CF^{(+)}_1\simeq&2(-)^{1-h\over t}(-\eta_{\rm ws})^{-{2\over t}}
{\Gamma(2+{2h-1\over t})\Gamma(-{2\over t})\over
\Gamma(1-{1\over t})\Gamma(1+{2h-2\over t})}\cr
}}
One can check that the power of $\eta_{\rm ws}$ is precisely correct
for describing the contribution of $\Psi_{-1}$. Plugging
\largeetaf\ into $a_1$ \lle, we find that this contribution
indeed vanishes for $|2h'-1|=1$ and is non-zero otherwise.

A similar check can be performed on the two point function $G_{t\over2}$
\llbt. The constraint \fhttwosim\ was obtained by analyzing the
contribution of the identity operator in the limit $z\to\bar z$.
We know from \ooe\ that the other boundary operators that can in principle
contribute to \llbt\ in this limit are the degenerate boundary operators
$\Psi_t$ and $\Psi_{1-t}$. However, an explicit calculation leads to
the conclusion that these contributions vanish (the outline of the proof
appears at the end of appendix B).

As we will see later, the above is the beginning of the emergence of
a structure somewhat similar to that found for Liouville branes in
\ZamolodchikovAH. The open string spectrum contains a finite number of
degenerate operators with half integer spin (of the form \nullrszbo).

At this point, one can in principle continue developing the bootstrap
approach by studying other correlation functions. For example, by
analyzing the two point function on the upper half plane of the
higher degenerate operators \nullrsz\ with $\Phi_h$,
$\langle\Phi_{(1-r)/2}\Phi_h\rangle$, one can compute the
amplitude for creating a degenerate operator $\Psi_{-n}$
with $n>1$ (and $n\in Z$) by taking a generic bulk operator
$\Phi_h$ to the boundary. As we explained earlier, we expect
to find this amplitude to be non-zero for the boundary state
corresponding to a certain $h'\in Z/2$, when $n\leq2|h'|$. It
would be interesting to verify this.

Another interesting problem is to analyze the two point function
of generic bulk operators on the upper half plane $\langle
\Phi_h\Phi_h\rangle$ and show that as the bulk operators approach
the boundary, they only create degenerate operators \nullrszbo.
We will not proceed further in this direction here; instead we
will turn next to the analysis of the annulus amplitude, which
adds insight about the theory.

\lref\AffleckGE{I.~Affleck,
``Conformal Field Theory Approach to the Kondo Effect,''
Acta Phys. Polon. B {\bf 26}, 1869 (1995) [cond-mat/9512099].}

\newsec{Modular bootstrap (I)}

The problem that we would like to address in this section
is the spectrum of open strings connecting two D-branes
corresponding to, say,
$|2h'-1|=n,m$ (or the spectrum of $(n,m)$ strings, in short).
A priori, the most naive thing to expect is that these strings
belong to the finite dimensional, degenerate representations
of the diagonal $SL(2)$ with half integer spin. More precisely,
one might expect that if we write $n=2j_1+1$, $m=2j_2+1$, then
the $(n,m)$ strings belong to finite dimensional representations
of $SL(2)$ which appear in the product\foot{Recall that the finite
dimensional spin $j$ representation corresponds to the degenerate
boundary operator $\Psi_{-j}$ \nullrszbo.}
\eqn\prodspins{{\bf j_1}\otimes {\bf j_2}=
|{\bf j_1}-{\bf j_2}|\oplus\cdots\oplus{\bf j_1}+{\bf j_2}.}

\noindent
Some of the reasons why this is a natural thing to expect are:
\item{(1)} Algebraically, since the boundary states were
found to be labeled by an integer $|2h'-1|$ corresponding to
a degenerate representation of $SL(2)$ with half-integer
spin, it is natural for the open strings to arrange themselves
in the representations that one obtains by sending these
degenerate operators to the boundary. Also, both the algebraic
analysis and the resulting structure are very reminiscent of
$SU(2)$ branes, which makes the spectrum \prodspins\ seem plausible.

\item{(2)} Recall the situation for $SU(2)$. There is a basic
Cardy state, which algebraically corresponds to $j=0$ in the sense
of \prodspins, and geometrically describes a $D0$-brane at a
point on the three-sphere. The open strings that live on this brane
belong to the current algebra block of the identity (in agreement with
\prodspins). Now consider the boundary state corresponding to
$n$ $D0$-branes on the sphere. Algebraically, it corresponds
to a tensor product of the basic Cardy state describing a single $D0$-brane
and $n\times n$ Chan-Paton factors; geometrically it describes
$n$ coincident $D0$-branes on the sphere. This boundary state is
marginally unstable: it contains a marginally relevant operator
$J^a(z)I_a$, $(a=3,+,-)$,
where $J^a(z)$ is the $SU(2)$ current which is conserved on the
upper half plane (evaluated at the boundary), while $I_a$ is a
constant $n\times n$ matrix representing $SU(2)$ in terms of the
Chan-Paton degrees of freedom. In the presence of this interaction,
the worldsheet CFT flows \AffleckGE\ to an interacting BCFT,
described by the $SU(2)$ boundary state corresponding to spin $(n-1)/2$.
Geometrically, the $n$ $D0$-branes expand into a finite size
two-sphere inside $S^3$. The situation is expected to be similar
in the $SL(2)$ case. The boundary state with $|2h'-1|=1$ corresponds
to a D-instanton in $AdS_3$ (we will motivate this further below), and
should have only the current block of the identity living on it.
The boundary state with $|2h'-1|=n$ that we constructed above corresponds
to $n$ D-instantons, presumably again in the presence of an interaction
of the form $J^a(z)I_a$. Thus, $(n,m)$ strings connect an
aggregate of $n$ instantons to $m$ instantons. One would
expect to find $nm$ $SL(2)$ primaries in this sector.
This is precisely what one finds if the operators belong to the
representations \prodspins.

\noindent
With the above comments as motivation, we will analyze in this
section the annulus partition sum corresponding to the degenerate
representations with half-integer spin, and its properties under
the modular transformation $\tau\to-1/\tau$.

We will be computing the character
\eqn\charsltwo{
Z(\tau, u)={\rm Tr}\; q^{L_0-{c\over 24}}e^{2\pi iuJ_3^0}, \qquad
q=e^{2\pi i\tau}~.}
We are really interested in the character with $u=0$ but
it might be convenient to compute with finite $u$ and take
$u$ to be small at the end of the calculation.

For a degenerate representation with spin $j={r-1\over2}$,
$m$ takes values in the finite range
$m=-j,-j+1,\cdots,j-1,j$.
Performing the trace \charsltwo\ one finds
\eqn\zrrr{Z_r(\tau,u)=
{\sin\pi ru\over\sin\pi u}
{q^{r^2\over 4t}\over
q^{1\over8}\prod_{n=1}^\infty(1-e^{2\pi iu}q^n)(1-q^n)
(1-e^{-2\pi iu}q^n)}~.}
This can be concisely written in terms of a theta function,
\eqn\thdef{\theta_1(u,\tau)=2q^{1/8}\sin\pi u
\prod_{n=1}^\infty(1-e^{2\pi iu}q^n)(1-q^n)
(1-e^{-2\pi iu}q^n)}
as
\eqn\zrtheta{Z_r(\tau,u)={2q^{r^2\over 4t}\sin\pi ru \over
\theta_1(u,\tau)}}
Now perform a modular transformation,
\eqn\modtrans{(\tau',u')=(-{1\over\tau}, {u\over\tau})~.}
The transformation property of the theta function is given
\eg\ in \ref\jpolch{J. Polchinski, ``String Theory,'' v. 1, p. 215,
Cambridge University Press (1998).}
\eqn\transftheta{\theta_1({u\over\tau},-{1\over\tau})=
-i(-i\tau)^{1/2}\exp(\pi iu^2/\tau)\theta_1(u,\tau)~.}
Thus, we have
\eqn\ztprupr{Z_r(\tau', u')={2i(q')^{r^2/4t}\sin\pi ru'\over
(-i\tau)^{1/2}\exp(i\pi u^2/\tau)\theta_1(u,\tau)}~.}
Using the fact that
\eqn\simpfy{2i(q')^{r^2/4t}e^{-i\pi u^2/\tau}\sin\pi ru'
=\exp\left[-{i\pi\over\tau}({r^2\over 2t}+u^2-ru)\right]
-\exp\left[-{i\pi\over\tau}({r^2\over 2t}+u^2+ru)\right]}
and
\eqn\zzident{\sqrt{i\over\tau}e^{-{\pi i\over\tau}(p')^2}=
\int_{-\infty}^\infty dp e^{\pi i\tau p^2}e^{2\pi ipp'}}
we conclude that:
\eqn\zrtransf{Z_r(\tau',u')={1\over\theta_1(u,\tau)}\int_{-\infty}^\infty
dp e^{\pi i\tau p^2}\left(
e^{2\pi ip\sqrt{{r^2\over 2t}+u^2-ru}}-
e^{2\pi ip\sqrt{{r^2\over 2t}+u^2+ru}}\right)~.}
We can now send $u\to 0$ and look at the resulting expressions.

The original degenerate character \zrrr\ becomes
\eqn\uzerozr{Z_r(\tau, u\to 0)={r q^{r^2/4t}\over \eta^3(\tau)}~.}
The modular transformed character \zrtransf\ goes to
\eqn\clstzr{Z_r(-{1\over\tau}, u'\to 0)={1\over\eta^3(\tau)}
\int_{-\infty}^\infty dp e^{\pi i\tau p^2}
e^{2\pi i{pr\over\sqrt{2t}}}(-ip\sqrt{2t})~.}
We note in passing that
\clstzr\ can be obtained directly from \uzerozr\ by using
\eqn\modeta{\eta(-{1\over\tau})=(-i\tau)^{1/2}\eta(\tau)}
and \zzident:
\eqn\aaiden{p'(-i\tau)^{-3/2}e^{-{\pi i\over\tau}(p')^2}=
{1\over 2\pi i}{d\over dp'}\sqrt{i\over\tau}e^{-{\pi i\over\tau}(p')^2}=
\int_{-\infty}^\infty dp p e^{\pi i\tau p^2}e^{2\pi ipp'}~.}

\noindent
The modular transformation \modtrans\ takes us from the open string
channel to the closed string channel. The original partition sum
\uzerozr\ can be thought of, for example, as the trace over open string states
stretched between a brane with $|2h'-1|=r$ and the brane with $|2h'-1|=1$
(\ie\ the partition sum of $(r,1)$ strings). The transformed partition sum
is a trace over closed strings that can be exchanged by the branes.

By comparing the closed string channel expression \clstzr\ to the
contribution of closed strings with spin $j$, one concludes that
the states exchanged by the branes belong to the principal continuous
series
\eqn\jlamb{j=-{1\over2}+i\lambda.}
The momentum in the radial direction, $\lambda$, is related to $p$ in \clstzr\
via the rescaling  $\half p^2=-\lambda^2/t$ or
\eqn\plambrel{p=\lambda\sqrt{-{2\over t}}.}
Using this rescaling, and remembering that $\lambda$ and $-\lambda$ \jlamb\
should be identified (this follows from \teschone), one can rewrite the
partition sum \clstzr\ as
\eqn\zzrrff{Z_r(-1/\tau)=-2\sqrt{-2/t}
\int_0^\infty d\lambda\chi_{\lambda}(q)
\sinh({2\over t}\pi\lambda r)}
where
\eqn\contlamb{\chi_\lambda(q)
={2\lambda q^{-\lambda^2/t}\over\eta^3(\tau)}.}
A potentially puzzling aspect of \contlamb\ is the apparent
absence of a contribution from the sum over eigenvalues of $J^3$,
which naively gives an infinite
multiplicative factor. We will not discuss this issue in detail
here, but would like to make a few remarks about it.

The origin of the problem is the fact that we are considering
here closed string exchange between coincident branes. In flat space,
this is usually dealt with by separating the branes in a transverse
direction, and computing the partition sum as a function of the
separation. This could also be done here. For example, consider the
annulus amplitude corresponding to $(r,1)$ strings, with one of the branes
located at ${\rm Im} x=0$ and the other at ${\rm Im} x= x_0$.
This can be implemented by imposing the symmetric boundary conditions
\symcur\ on one of the boundaries, and boundary conditions twisted by
an automorphism (see section 7) on the other boundary.
This would regularize the divergence in question.

When the separation between the branes goes to zero, one expects,
as in flat space, to find that the partition sum in the closed
string channel gets additional factors of $\tau$, which are the remnant
of the divergence. This is indeed what happens here, but as explained
in eqs. \clstzr\ -- \aaiden, we chose to trade the powers of $\tau$
for a power of $p$ (or $\lambda$). Thus the factor of $\lambda$ in
\contlamb\ is the remnant of the sum over $m$.

Returning to \zzrrff, general properties of BCFT imply that we should
be able to write the partition sum of $(r,1)$ strings as
\eqn\zrone{Z_{r,1}(q')=\int d\lambda \chi_\lambda(q)\Psi_r(\lambda)
\Psi_1(-\lambda)}
where the wavefunction $\Psi_r(\lambda)$ is proportional to
the one point function of the bulk primary with $h=\half+i\lambda$
in the state labeled by $r$. The wavefunction in \zzrrff\ is proportional
to $\sinh(2\pi\lambda r/t)$. Comparing to the one point function \solnfh\
taking into account the map between the variables $h,h'$ in \solnfh\ and
$\lambda,r$ here,
$$h=\half+i\lambda;\;\;r=|2h'-1|$$
we see that the $h$ dependence is correct. Comparing \zzrrff\ and \zrone\
we furthermore find that
\eqn\psrone{\Psi_r(\lambda)\Psi_1(-\lambda)
=-2\left(-{2\over t}\right)^{\half}\sinh({2\over t}\pi\lambda r)~.}
In particular,
\eqn\psionel{\Psi_1(\lambda)\Psi_1(-\lambda)
=-2\left(-{2\over t}\right)^{\half}\sinh({2\over t}\pi\lambda)~.}
A solution to this is
\eqn\psiosol{\Psi_1(\lambda)=\left(-{t\over2}\right)^{1\over 4}
{\sqrt{2\pi/\lambda}\over\Gamma(-{2\over t}i\lambda)}~.}
In general, we should be able to write the partition sum of $(r,r')$
strings as
\eqn\znmexp{Z_{r,r'}(q')=\int d\lambda \chi_\lambda(q)\Psi_r(\lambda)
\Psi_{r'}(-\lambda)~.}
Plugging \psrone, \psiosol\ into \znmexp\ and using the identity
\eqn\ssss{\sum_{n=0}^{\min(r,r')-1}\sinh({2\pi\lambda\over t})
\sinh({2\pi\lambda\over t}(r+r'-2n-1))
=\sinh({2\pi\lambda\over t}r)\sinh({2\pi\lambda\over t}r'), \qquad r,r'\in Z}
we find that
\eqn\zrrzr{Z_{r,r'}(q)=\sum_{n=0}^{\min(r,r')-1}Z_{r+r'-2n-1}(q),
\qquad r=2j+1,\,\, r'=2j'+1, \qquad j,j'=0,{1\over 2},1,\dots}
This is precisely the spectrum proposed in the beginning of this
section in \prodspins.

To summarize, starting with the assumption that $(r,1)$ strings transform
in the degenerate spin $(r-1)/2$ representation, we showed that the
spectrum of $(r,r')$ strings is given by \prodspins\ and that the
wavefunction of the boundary state labeled by $r$ is proportional to
the one point function $f(h)$ \solnfh, in agreement with expectations.
This provides strong evidence for the validity of the overall picture.

\newsec{$SL(2)$ conformal field theory on the upper half plane (II)}

In the previous sections we have described D-branes that correspond to
instantons in $AdS_3$. One certainly expects to find branes that
correspond to real, physical objects in $AdS_3$ as well. In this section
we will outline the algebraic structure that underlies a class of branes
that correspond (as we will see later) to Euclidean and Lorentzian
two dimensional worldvolumes embedded in $AdS_3$.

Algebraically, the branes that we will construct correspond to the principal
discrete series $\half<h\in \IR$, and the continuous series $h=\half+i\lambda$,
$0<\lambda\in \IR$ (recall that the instantonic branes of the previous
sections correspond to finite dimensional degenerate representations of
$SL(2)$, \nullrsz). Thus, they have a much richer spectrum of boundary
$SL(2)$ primaries $\Psi_h(x;z)$ living on them than the branes described
in sections 3,4. This is natural from the geometric point of view, since
these branes are extended in two directions.

A fact that will play a role in the discussion is that one can turn on
a constant electric field\foot{We use the terminology suitable for $1+1$
Minkowski worldvolume, although some of the branes we will construct are
Euclidean.} on these D-branes without breaking worldsheet conformal
invariance (or going off-shell in spacetime). In flat spacetime, this
is described by the Born-Infeld action
\eqn\lbi{\CL_{BI}={1\over g_s}\sqrt{1-E^2}}
and one expects to find a similar structure
for branes in $AdS_3$ \BachasFR.

In order to turn on an electric field on the worldvolume, one needs to add
to the worldsheet Lagrangian the vertex operator of a zero momentum photon.
The relevant vertex operator is
\eqn\iiibndry{I_b=\int dz J(x;z)\Psi_1(x;z).}
The integral over $z$ runs over the real line, the boundary
of the worldsheet. The real variable $x$ parametrizes the boundary of the
spacetime upper half plane. $\Psi_1$ is a primary of $SL(2)$ which lives on
the boundary; it has worldsheet dimension zero \bounddim, and spacetime
dimension one. $J(x;z)$ is the diagonal (conserved) $SL(2)$ current, restricted
to the boundary.

Many of the properties of $I_b$ can be understood by noting its similarity
to the vertex operator of the zero momentum dilaton,
\eqn\iii{I=-{1\over k^2}\int d^2z J(x;z)\bar J(\bar x;\bar z)\Phi_1(x,\bar x;
z,\bar z),}
introduced and discussed in \ks\ (see also \gksch).
$I_b$ is a marginal operator on the worldsheet; adding it to the action,
\eqn\sepert{\SS=\SS_0-E\int dz J(x;z)\Psi_1(x;z),}
corresponds to turning on an electric field on the brane (which,
as mentioned above, is a modulus). While naively $I_b$ depends on
$x$, one can show, along the lines of \ks, that in fact
\eqn\xib{{\partial I_b\over\partial x}=0,}
in agreement with the fact that its spacetime scaling dimension is zero
\ks, and that it describes a {\it constant} electric field.

As discussed in \gksch, $I$ is a dimension zero operator which is not
proportional to the identity operator, despite the fact that it satisfies
$\partial_x I=\partial_{\bar x}I=0$ (\ie\ it is constant in correlation
functions, but not necessarily the same constant in different correlation
functions). Similarly, $I_b$ is a dimension zero boundary operator in the
spacetime CFT, which is not proportional to the identity (as we will see).
Just like $I$ keeps track of sectors with different central charges \gksch,
$I_b$ keeps track of sectors with different boundary conditions on the
upper half plane ${\rm Im} x\ge 0$.

Since the operator $I$ corresponds to the zero momentum dilaton, adding it
to the worldsheet action changes the dilaton expectation value and thus the
closed string coupling \refs{\ks,\gksch}. Similarly, adding $I_b$ to the
worldsheet action \sepert\ changes the electric field on the brane and thus
the open string coupling. As discussed in \refs{\gks,\ks}, the expectation
value of $I$ is arbitrary classically but is
quantized in the full quantum theory. Similarly, the expectation
value of $I_b$ is classically arbitrary, but quantum mechanically
it is quantized\foot{This quantization is associated with the
fact that an electric field on a D-string is equivalent to fundamental
string charge, and simply corresponds to bound states of one D-string
and $n$ fundamental strings. The fact that $n$ is integer quantizes
the electric field.} (this is standard for the gauge field on D-strings
in flat space, and is explained in the context of $AdS_3$ branes
in \BachasFR).

For some calculations it is useful to note that $I_b$ is closely related
to the ``boundary Wakimoto screening operator.''
In the bulk case, this is shown in \gksch.
Similar arguments lead to the conclusion that in ``Wakimoto variables''
$(\phi,\beta,\gamma)$, adding $I_b$ to the action as in \sepert\ is
equivalent to perturbing by
\eqn\sebeta{\SS=\SS_0-E\int dz\beta e^{-\half Q\phi(z)}.}
We would like next to repeat the analysis of section 3 for the extended
branes, and derive the analog of eq. \fsimple\ for the one point function
\llaa. Thus, consider the two point function $G_{-\half}(h)$ \lla.
We can again calculate the contribution of the identity in the limit
$z\to\bar z$ in two different ways, thereby obtaining a constraint
on $U(h)$. The discussion leading to eq. \finf\ is unchanged, since it
does not depend on any detailed properties of the boundary state in which
the calculation is done. However, the factorization property \ppuu\ is
no longer valid here \refs{\FateevIK,\ZamolodchikovAH}.
The reason is that there is now a
continuum of boundary operators that can be created as $\Phi_{-\half}\to$
boundary, and the ``contribution of the identity'' in \lla\ as $z\to\bar z$
is a coherent effect of this continuum.

Nevertheless, we can proceed in a different way since, in analogy to
\refs{\FateevIK,\gksch}, the physics associated with $\Phi_{-\half}\to$
boundary is expected to be perturbative in $E$ \sebeta\ for branes which
reach the boundary of $AdS_3$. We can expand the action to first order
in $E$ and use free field OPE's to deduce the structure constant in \llg,
\eqn\stao{\lim_{w\to\bar w}\Phi_{-\half}(x,\bar x;w,\bar w)=A_0{x-\bar x\over
(w-\bar w)^{2\Delta_{-\half}}}+\cdots.}
Using \modehalf\ we find that
\eqn\Aoform{{A_0\over(w-\bar w)^{2\Delta_{-\half}}}=E\int_{-\infty}^\infty
dz\beta e^{-\half Q\phi(z)} V_{-\frac32;-\half,\half}(w)~.}
As explained in \gksch, eq. (4.13), we have
\eqn\vmhf{V_{-\frac32;-\half,\half}(w)=-{2\over\pi}\RR(-\half)\gamma(w)
e^{\half Q\phi(z)}~.}
Plugging \vmhf\ in \Aoform\ and performing the integral using free
field contractions, we find
\eqn\aostc{A_0=\tilde C(t) E,}
where $\tilde C(t)$ is a calculable constant (function of $t$ \defttt)
that we will not write explicitly.

Returning to \ppuu, we thus have in this case
\eqn\pexten{
\langle\Phi_{-\half}(x,\bar x;z,\bar z)\Phi_h(y,\bar y;w,\bar w)\rangle
\simeq{(w-\bar w)^{-2\Delta_h}\over
(z-\bar z)^{2\Delta_{-\half}}}{x-\bar x\over(y-\bar y)^{2h}}
\tilde C(t) EU(h)~.}
We can now compare \pexten\ to \finf\ (as in section 3) and conclude
that the rescaled one point function $f(h)$ \fhdef\ satisfies
\eqn\nowgive{C(t)Ef(h)=f(h-{1\over 2})+f(h+{1\over 2})}
where $C(t)$ is another calculable constant, whose precise form will
not be needed here. In addition, $f(h)$ still satisfies the reflection
property \fhsymm. The solution of this equation that is relevant here has
the form (compare to \solnfh)
\eqn\ffhhaa{f(h)=A(t)\sin[{\pi\over t}(2h-1)(2h'-1)].}
A few remarks are in order:
\item{1.}
Unlike \solnfh, in \ffhhaa\ we are computing
$\langle\Phi_h\rangle$ {\it without} dividing by the partition sum
$\langle 1\rangle$ on the upper half plane.
\item{2.}
In the next section we will verify \ffhhaa\ by an analysis of the annulus
amplitude, but for now we will assume the form \ffhhaa\ and study some
of its properties.
\item{3.}
Note that the equation \nowgive\ does not determine the overall normalization
of $f(h)$. This ambiguity is absorbed in the factor $A(t)$ in \ffhhaa.

\noindent
Like in section 3, one can think of $h'$ as labeling the representation
to which the boundary state in question corresponds.
Equation \nowgive\ relates
$h'$ to the electric field by:
\eqn\lrebcc{\cos{\pi\over t}(2h'-1)=\half E C(t).}
This is an interesting relation: for $h'$ belonging to the
principal discrete series ($h'\in \IR$), the electric field
is bounded from above,
\eqn\lbnd{|E|\leq {2\over C(t)}.}
For the continuous series, $h'=\half+i\lambda$, the electric field
is instead bounded from below,
\eqn\ubnd{|E|\geq {2\over C(t)}.}
This is reasonable, since the principal discrete series is associated
with massive particles in $AdS_3$, while the continuous series describes
tachyons, and it is well known that above a certain critical field,
D-branes become tachyonic (as is obvious \eg\ from \lbi). This
interpretation suggests that the bound in \lbnd, \ubnd\ is the
critical electric field of open string theory.

To see that this is indeed the case, consider the partition sum
on the disk with the boundary conditions corresponding to a certain
boundary state labeled by $h'$. It is expected to depend on the electric
field via the Born-Infeld form \lbi. Our formalism allows one to compute
the partition sum as follows. First note that if we know the normalization
factor $A(t)$ in \ffhhaa, we can compute the partition sum by setting
$h=0$ in \ffhhaa. Let us assume for now (we will return to this assumption
shortly) that $A(t)$ does not
depend on the boundary state (\ie\ on $h'$), as implied by the notation.
Then, ignoring a $t$ dependent overall factor, we have
\eqn\zdisk{Z_{\rm disk}\simeq\sin{\pi\over t}(2h'-1).}
Using \lrebcc\ to write this in terms of $E$, we find
\eqn\zbif{Z_{\rm disk}\simeq\sqrt{1-\left(\half E C(t)\right)^2}.}
After rescaling $E$, this is precisely of the Born-Infeld form \lbi,
and the critical field is indeed the one we deduced from \lbnd, \ubnd\
before.

Thus, the branes corresponding to $h'$ in the principal continuous
series are tachyonic, while those in the principal discrete series are
massive, in agreement with what one would expect.

We see that it is very natural to expect that $A(t)$ in \ffhhaa\ is
independent of $h'$,
but is this really the case? We believe the
answer is yes, but will leave a detailed verification of this to
future work. We would like to make some comments on this matter:

\item{(1)}
$A(t)$ can be computed by using an idea from \FateevIK.
The basic point is the following: we know the $h$ dependence of $f(h)$
-- it is given by \ffhhaa. At the same time, there is an infinite number
of values of $h$ for which \ffhhaa\ can be
computed using free field methods, by perturbing in the Wakimoto coupling
(the coefficient of $I$ \iii\ in the action -- see \gksch) and in $E$.
This calculation can be used to determine $A(t)$.

\item{(2)}
It is natural to expect that the only dependence on the
electric field $E$ in \ffhhaa\ is via its relation to $h'$, \lrebcc;
this was found to be the case in a similar calculation in boundary
Liouville theory in \FateevIK.

\noindent
Equation \ffhhaa\ was found above to be valid for the principal discrete
and continuous series. It is interesting to note that it is valid for the
finite dimensional representations discussed in section 3 as well. To show
this, one computes the partition sum on the upper half plane, by setting
$h=0$ in \ffhhaa\ recalling the finite renormalizations \fhdef, \rhottt,
\uhone. Dividing $f(h)$ \ffhhaa\ by the resulting partition sum, one finds
precisely the form \solnfh.

The last comment also suggests that it might be possible to interpolate
between the two dimensional branes of this section and the pointlike
branes discussed in sections 3,4, in analogy to the situation in flat
space where one can turn D2-branes into (collections of)
D0-branes by turning on a large magnetic field on the worldvolume
of the D2-brane.

\newsec{Modular bootstrap (II)}

As in the analysis of the pointlike branes before, we can supplement the
bootstrap analysis with information from the annulus. Consider the annulus
partition sum for open strings stretched between a brane labeled by
$h'$ (from the principal discrete or continuous series) and the D-instanton
brane corresponding to $h'=0$, or $r=1$ in sections 3,4. It is natural to
expect that these open strings belong to the $SL(2)$ representation labeled
by $h'$. Thus, the partition sum is given by
\eqn\zhpq{Z_{h',1}=\chi_{h'}(q),}
where $\chi_{h'}$ is the open string character corresponding to
the representation $h'$,
\eqn\chihhh{\chi_{h'}=
{2\lambda' q^{{1\over4t}(2h'-1)^2}\over\eta^3(\tau)},}
where, as before, $\lambda'$ is related to $h'$ via
\eqn\hplam{h'=\half+i\lambda'.}
$\lambda'$ is real for the continuous representations and imaginary
for the discrete ones. We can now perform the modular transform
$\tau'=-1/\tau$, using \zzident, \modeta. This gives
\eqn\modtadst{Z_{h',1}(q')
=-2\sqrt{-{2\over t}}\int_0^\infty d\lambda
\chi_{\lambda}(q)\sin{4\pi\lambda\lambda'\over t}}
The same logic that led to \znmexp\ leads to the conclusion
that the wave function $\Psi_{h'}(\lambda)$ corresponding
to the boundary state labeled by $\lambda'$ is given by
\eqn\wavefadst{\Psi_{h'}(\lambda)\Psi_1(-\lambda)
=-2\sqrt{-{2\over t}}\sin{4\pi\lambda\lambda'\over t}}
Using $2i\lambda=2h-1$,
we conclude that the one point function $f(h)$
in the boundary state labeled by $h'$ is proportional to
\eqn\ffhhbb{f(h)\propto\sin{\pi\over t}(2h-1)(2h'-1)}
in agreement with the discussion of the previous section (compare
\ffhhbb\ to \ffhhaa).

It is also possible to find the spectrum of open strings stretched between
a brane labeled by $h'$ and an excited (multi-instanton) brane,
corresponding to the degenerate representation with $|2h'-1|=r$.
By using the expression
\eqn\zhpr{Z_{h',r}(q')=\int d\lambda\chi_\lambda(q)\Psi_{h'}(\lambda)
\Psi_r(-\lambda)}
and the wavefunctions \psrone, \wavefadst, we find
\eqn\zzzhhhrr{Z_{h',r}(q)=\sum_{m=0}^{r-1}\chi_{h'-{r-1\over2}+m}(q)}
from which we can read off the open string representations that appear
in the sector $(h',r)$.

Finally, one can also study the spectrum of open strings that
stretch between two branes labeled by $h_1'$ and $h_2'$.
This spectrum is expected to be similar to the familiar one
from closed string theory on $AdS_3$, with states belonging to the
principal discrete and continous series, and long string sectors
corresponding to twisted representations, as in \refs{\maloog,\ags}.
A detailed study of this is left for future work.


\newsec{Asymmetric gluing}

Up to this point, our discussion involved branes that are obtained by
imposing symmetric gluing conditions on the currents, \symcur.
As is well known, one can also study branes which preserve a different
diagonal $SL(2)$, related to the original one by the chiral application of
an automorphism of the algebra (say, just on the left
movers).~\foot{For instance, branes coresponding to a boundary
at $\bar x -x=c$, where $c$ is a constant, are obtained by imposing the
gluing conditions $J^a=\bar J^a$ with $J^a$ acting as the
differential operators on the r.h.s. of \rrr, with $x\to x+c$.}
Examples of automorphisms of the $SL(2)$ algebra \opealg\ are
\eqn\inneraut{
\eqalign{
J^\pm & \to- J^\mp;\;\;J^3\to- J^3\cr
J^\pm & \to- J^\pm;\;\;J^3\to J^3\cr
J^\pm & \to J^\mp;\;\;J^3\to- J^3~.\cr}}
Twisting with the first line of \inneraut, one finds
branes that correspond to the following boundary conditions on the
$SL(2)$ currents at $z=\bar z$:
\eqn\twistbnd{J^\pm(z)=-\bar J^\mp(\bar z);\;\;J^3(z)=-\bar J^3(\bar z).}
Thus, the $x$ dependence of the one point function \llaa\ changes.
For example, in the particular case \twistbnd, imposing the fact that the one
point function should satisfy $J^3-\bar J^3=0$ and $J^\pm-\bar J^\mp=0$
(using \rrr), one finds that
\eqn\lltwist{\langle\Phi_h(x,\bar x;z,\bar z)\rangle
={U(h)\over(1-x\bar x)^{2h}(z-\bar z)^{2\Delta_h}}~.}
Replacing \llaa\ by \lltwist\ one can now proceed with the bootstrap
discussion of the previous sections, and find the same set of branes
as before. The only difference is that this time the branes correspond
to a boundary at $x\bar x=1$, the boundary of the unit disk (in Euclidean
space), which is
conformally related to the upper half plane that appeared in the
construction of the earlier sections. Similarly, the second line
of \inneraut\ gives rise to a boundary at $x+\bar x=0$ in the spacetime
CFT, while the third line corresponds to a boundary at $x\bar x+1=0$.
The latter has no solutions in Euclidean spacetime, but makes sense in
Minkowski space.

To summarize, as one would expect, twisted boundary conditions on the currents,
like \twistbnd, give rise to the same set of branes that was discussed above.
The branes give rise to a boundary in the spacetime CFT, and for different
diagonal $SL(2)$ algebras one finds boundaries oriented in a different way in
$x$ space. Of course, in Minkowski space, some of the constructions give rise
to timelike boundaries while others produce spacelike ones, but from the
point of view of the algebraic analysis performed here, this does not
influence most of the results.

\newsec{Discussion}

In this paper we have performed an algebraic
analysis of D-branes in $AdS_3$.
We solved the bootstrap equations for the one point
function of bulk operators \llaa,
which are the wavefunctions of the boundary states.
We also studied the annulus
partition sum, which provided additional
information on the spectrum of open strings
stretched between different branes.
We found three classes of branes, labeled by representations of $SL(2,R)$:
\item{(1)} Finite dimensional degenerate representations with
\eqn\hprntwo{h'=-{n\over2},\qquad n=0,1,2,3,\cdots.}
\item{(2)} Principal continuous series,
\eqn\hprlamb{h'=\half+i\lambda,\qquad \lambda\in\IR.}
\item{(3)} Principal discrete series,
\eqn\hprdiscr{\half<h'\in\IR.}

\noindent
It is natural to ask what do the different branes correspond to geometrically.
We argued that the first class of branes \hprntwo\ are zero dimensional,
while the last two \hprlamb, \hprdiscr\ are two dimensional. We showed that
the continuous series branes \hprlamb\ are tachyonic, since they have a
supercritical electric field on the worldvolume, while the discrete series
branes \hprdiscr\ have a subcritical electric field and positive mass squared.

The authors of \BachasFR\ analyzed the
semiclassical geometry of branes in $AdS_3$
by studying the corresponding (twined) conjugacy classes.
The regular conjugacy classes
are characterized by ${\rm Tr} g$, where $g$ is the $SL(2)$ matrix \defg:
\eqn\gofsl{g={1\over l}\left(\matrix{x_0-x_1 & x_2-x_3\cr
                            x_2+x_3 & x_0+x_1\cr}\right),}
The conjugacy classes are:
\item{(1)} $g=1$, which is the point $x_1=x_2=x_3=0$, $x_0=l$ on $AdS_3$.
\item{(2)} Two dimensional de Sitter space ($dS_2$), corresponding to
\eqn\trgtc{{\rm Tr} g=2C,}
with $|C|>1$.
\item{(3)} Two dimensional hyperbolic plane ($H_2$),
corresponding to \trgtc\ with
$|C|<1$.
\item{(4)} Light-cone, corresponding to $|C|=1$ in \trgtc.

\noindent
It was also shown in \BachasFR\ that $dS_2$ branes have a supercritical
electric field on their worldvolume, and are thus tachyonic, while the
Euclidean $H_2$ branes have a subcritical electric field. The $H_2$ branes
also have a Minkowski counterpart ($AdS_2$ branes), obtained by considering
twined conjugacy classes, \ie\ twisting the gluing conditions
as in section 7.

There is a natural map between the branes constructed here and those
of \BachasFR. The basic brane, corresponding to $h'=0$ \hprntwo, is associated
with the conjugacy class $g=1$. The multi-instanton branes \hprntwo\ with
$n>1$ correspond to $n$ of the $g=1$ branes, in some sort of extended
bound states. It is not quite clear to us what is the semiclassical description
of these states\foot{This is similar to the situation with excited branes in
\ZamolodchikovAH, which were constructed algebraically, but are not understood
semiclassically.}.

The $dS_2$ branes of \BachasFR\ correspond to the continuous series \hprlamb,
while the $H_2$ and $AdS_2$ branes correspond to the principal discrete series
\hprdiscr. As explained in section 7, $H_2$ and $AdS_2$ branes give rise to
equivalent boundary states, with the boundary oriented in $x$
space in a different way in the two cases.

The role of the light-cone conjugacy class, $|C|=1$ in \trgtc, is less clear.
It is possible that the $n>1$ branes \hprntwo\ expand into the light-cone, but
this requires a better understanding.

\lref\BalasubramanianSN{
V.~Balasubramanian, P.~Kraus and A.~E.~Lawrence,
``Bulk vs. boundary dynamics in anti-de Sitter spacetime,''
Phys.\ Rev.\ D {\bf 59}, 046003 (1999)
[hep-th/9805171].
}
\lref\BanksNR{
T.~Banks and M.~B.~Green,
``Non-perturbative effects in AdS(5) x S**5 string
theory and d = 4 SUSY  Yang-Mills,''
JHEP {\bf 9805}, 002 (1998)
[hep-th/9804170].
}
\lref\BalasubramanianDE{
V.~Balasubramanian, P.~Kraus, A.~E.~Lawrence and S.~P.~Trivedi,
``Holographic probes of anti-de Sitter space-times,''
Phys.\ Rev.\ D {\bf 59}, 104021 (1999)
[hep-th/9808017].}
\lref\KarchFR{
A.~Karch and L.~Randall,
``Localized gravity in string theory,''
[hep-th/0105108].}
\lref\kara{
A.~Karch and L.~Randall,
``Open and closed string interpretation of SUSY CFT's on branes with
boundaries,''
[hep-th/0105132].}
\lref\GiveonJG{
A.~Giveon and M.~Rocek,
``Supersymmetric string vacua on AdS(3) x N,''
JHEP {\bf 9904}, 019 (1999)
[hep-th/9904024].}
\lref\sw{N.~Seiberg and E.~Witten,
``The D1/D5 system and singular CFT,''
JHEP {\bf 9904}, 017 (1999)
[hep-th/9903224].}
\lref\PetropoulosQU{
P.~M.~Petropoulos and S.~Ribault,
``Some remarks on anti-de Sitter D-branes,''
hep-th/0105252.
}

We finish with a few remarks about the above identification.
Some of the comments below require much more work:
\item{(1)}
Naively, it seems that there is a disagreement between the geometry
of the branes in \BachasFR\ and here.
The conjugacy classes \trgtc\ correspond
near the boundary of $AdS_3$
($u\to\infty$ in Poincare coordinates \metricupypm, \defg) to
\eqn\ggpone{\gamma\bar\gamma+1=0.}
On the other hand, we saw in section 3
that symmetric gluing \symcur\ leads to
branes localized at $x=\bar x$.
Since near the boundary of $AdS_3$, $\gamma\simeq x$,
here one finds a boundary at
\eqn\xbarx{\gamma=\bar\gamma.}
The discrepancy between \ggpone\ and \xbarx\ seems to be due to a different
definition of ``diagonal gluing'' in the two cases. As we saw in section 7,
one can get a boundary at $x\bar x+1=0$ by preserving a different $SL(2)$ then
\symcur. Presumably, this is what corresponds to the untwisted construction in
\BachasFR.
\item{(2)}
We argued above that string theory on $AdS_3$ in the presence of
D-branes corresponds to studying the spacetime CFT on the upper half
plane ${\rm Im}x\ge 0$. This might seem puzzling, since the branes look
more like defects stretched along the real line \xbarx\ and it is not clear
why one should not think of the spacetime CFT as still living on the
whole complex plane, in the presence of a line defect corresponding to the
brane. Since the worldsheet CFT on $AdS_3$ is left-right symmetric, the
structure that we find in the presence of the branes appears to be
{\it consistent} with focusing on the theory on just the upper (or just the
lower) half plane. For left-right asymmetric
worldsheet theories one in general
expects problems with defining D-branes.
\item{(3)}
Another interesting question concerns the boundary conditions imposed
on the spacetime CFT by the introduction of the branes. The D-instanton
$h'=0$ \hprntwo\ seems to correspond to the basic Cardy
state in the spacetime CFT. The fact that from the worldsheet point
of view only the identity and its $\widehat{SL(2)}$ descendants live
on the brane implies, using the results of \refs{\gks,\ks}, that
in spacetime the (boundary) excitations of this brane include only
the identity and its Virasoro descendants. It is less clear what boundary
boundary CFT corresponds to the spectrum we found on the higher ($|h'|>0$)
multi-instanton branes, as well as the extended
$dS_2$, $AdS_2$ and $H_2$ branes.
It is an interesting remaining problem
to describe the corresponding BCFT directly in spacetime.
It seems that one must add boundary interactions of some sort;
their exact nature is left for future work.
\item{(4)}
One might be puzzled by the fact that the localized branes influence
the physics near the boundary at all. For example, the D-instanton, which is
associated with the conjugacy class $g=1$, is classically located at $r=0$ (or
$u=1$) in the coordinates \metads, \metricupypm, deep in the bulk of $AdS_3$.
How can it introduce a boundary in $x$ space, which naively parametrizes the
large $r$ part of space? The answer seems to be the following. In general in
anti-de-Sitter space, observables in the spacetime CFT (local operators)
correspond to wavefunctions that are supported near
the boundary of $AdS$ space. States in the spacetime CFT correspond to
normalizables wavefunctions that are typically supported at $u\simeq 1$.
Interactions take place in the bulk of the space, despite the fact that
the observables are defined at large $r$. The D-instanton, and more
generally the zero dimensional branes \hprntwo, do not introduce any new
{\it observables} into the theory. They do introduce an object, off which
the non-normalizable observables from the closed string sector can scatter,
and therefore it is natural that they have a
description of the sort proposed here.
\item{(5)}
A related puzzling aspect of the pointlike D-instantons is the
fact that they induce a {\it finite} size boundary in the spacetime
CFT -- a one dimensional boundary at, say, $x\bar x=-1$ in the
construction of \BachasFR. Naively one might expect that
they should give rise to a pointlike defect (or local insertion
in the spacetime CFT) at $\gamma=\bar\gamma=0$. This ``blow up'' of
the instantons seems to be related to a familiar phenomenon in the
context of the $AdS_5/SYM_4$ correspondence. It is known
\refs{\BanksNR,\BalasubramanianSN} that the radial location
of D-instantons in $AdS_5$ is related to the size of the
corresponding instantons in super Yang-Mills. D-instantons near
the boundary of $AdS_5$ correspond to small instantons, and as they
approach the horizon of $AdS_5$, they grow in SYM.
Similarly here, D-instantons located at $u=1$ (far from both the
horizon and the boundary of $AdS_3$) seem to correspond to objects
of size ``one'' in the spacetime CFT. What is perhaps surprising is that
closed string fields see the expanded instantons as sharply defined
objects (as signaled by the singularities of the closed string
correlators near the boundary in spacetime) -- one might have
expected an everywhere smooth behavior.
\item{(6)}
There are many possible extensions of the work described here.
It would be interesting to complete the bootstrap program and verify its
full consistency. Also, we have restricted attention to the ``short string''
sector of the model. Perhaps additional insight could be obtained by applying
our methods to sectors with long strings. It would
also be interesting to extend
the analysis to other related backgrounds, such as the Euclidean cigar,
$SL(2)/U(1)$. In that case it is known that one should find
both branes that are extended along the radial direction of the cigar,
and branes localized near the tip \refs{\ElitzurPQ,\adg}. It is natural
to expect that branes localized near the tip (which are of interest for
various application -- see section 1) will arise from the localized branes
in $AdS_3$, and the multi-instanton branes discussed here will correspond to
small disks localized near the tip.
\item{(7)}
It would also be interesting to understand the Liouville branes
\refs{\FateevIK,\TeschnerMD,\ZamolodchikovAH} from the perspective of
$SL(2)$ branes, by dividing by the Borel subgroup. One way of making a direct
connection is to consider string theory on $AdS_3$ in static gauge,
$\gamma=z$, $\bar\gamma=\bar z$. This corresponds to studying the system
in a vacuum containing a long string \refs{\gks,\sw}.
The transverse fluctuations
of such a string are described by the Liouville Lagrangian for the radial
coordinate $\phi$ \defphi\ \sw. D-branes which reach the boundary of $AdS_3$
intersect the worldsheet of the long string, and thus it is natural that
they give rise to worldsheet boundaries at a finite distance from the bulk,
as found for the extended branes in \FateevIK. Thus, these branes correspond
to the principal series solutions \hprlamb, \hprdiscr, and the boundary
cosmological constant $\mu_B$ in \FateevIK\ is a remnant of the electric field
$E$ on these branes. D-instantons do not reach the boundary of $AdS_3$,
and thus do not intersect the worldsheet of the long string. They give rise
in the Liouville description to a boundary at an
infinite distance from the bulk,
as found in \ZamolodchikovAH. It would be interesting to make this picture
more precise.
\item{(8)}
It would also be interesting to extend the present work
to study the properties of D-branes in supersymmetric string theories on
$AdS_3\times\NN$ \refs{\GiveonJG,\ags}. For example, the type II superstring
on $AdS_3\times S^3\times T^4$ has supersymmetric branes with
an $AdS_2\times S^2$ worldvolume (which were studied geometrically
in \refs{\BachasFR,\PetropoulosQU}).
Such supersymmetric $AdS_d$ branes in $AdS_{d+1}$
have some interesting properties, suggested recently in
\refs{\KarchFR,\kara}, and our techniques might be useful for
studying them.

\bigskip
\noindent{\bf Acknowledgements:}
We thank O. Aharony, C. Bachas, M. Berkooz, N. Itzhaki, J. Maldacena,
A. Parnachev, D. Sahakyan, N. Seiberg and A. B. Zamolodchikov for
discussions. This work is supported in part by the Israel Academy of
Sciences and Humanities -- Centers of Excellence Program,
the BSF -- American-Israel Bi-National Science Foundation,
the German-Israel Bi-National Science Foundation,
and the European RTN network HPRN-CT-2000-00122.
A.G. thanks the Einstein Center at the Weizmann
Institute for partial support. The work of D.K. is supported in part
by DOE grant \#DE-FG02-90ER40560. D.K. thanks the Weizmann Institute
for hospitality during the course of this work.

\appendix{A}{Some useful formulae}

The hypergeometric function is defined by the differential
equation for a function $u(x)$
\eqn\AAA{x(1-x)u''+\left[\gamma-(\alpha+\beta+1)x\right]u'-\alpha
\beta u=0}
This equation has two solutions:
\eqn\AAB{\eqalign{
u_1=&F(\alpha,\beta;\gamma;x)\cr
u_2=&x^{1-\gamma}F(\alpha-\gamma+1,\beta-\gamma+1;2-\gamma;x)\cr
}}
For small $x$, $F$ can be expanded as follows:
\eqn\AAC{F(\alpha,\beta;\gamma;x)=
\sum_{n=0}^{\infty}{(\alpha)_n(\beta)_n\over n!(\gamma)_n}x^n
=1+{\alpha\beta\over\gamma}x+
{\alpha(\alpha+1)\beta(\beta+1)\over2\gamma(\gamma+1)}x^2+\cdots~,}
where
\eqn\AACA{(a)_n\equiv a(a+1)\cdots (a+n-1)={\Gamma(a+n)\over\Gamma(a)}~.}
Two other identities that are sometimes useful are:
\eqn\AAD{\eqalign{
&F(\alpha,\beta;\gamma;x)=(1-x)^{\gamma-\alpha-\beta}
F(\gamma-\alpha,\gamma-\beta; \gamma;x)\cr
&{\partial F\over\partial x}(\alpha,\beta;\gamma;x)={\alpha\beta\over\gamma}
F(\alpha+1,\beta+1;\gamma+1;x)\cr
}}
Under $x\to 1/x$:
\eqn\AAE{\eqalign{
&F(\alpha,\beta;\gamma;x)=
{\Gamma(\gamma)\Gamma(\beta-\alpha)\over\Gamma(\beta)\Gamma(\gamma-\alpha)}
(-{1\over x})^\alpha F(\alpha,\alpha+1-\gamma;\alpha+1-\beta;{1\over x})
+\cr
&{\Gamma(\gamma)\Gamma(\alpha-\beta)\over\Gamma(\alpha)\Gamma(\gamma-\beta)}
(-{1\over x})^\beta F(\beta,\beta+1-\gamma;\beta+1-\alpha;{1\over x})
\cr}}
Under $x\to 1-x$:
\eqn\AAF{\eqalign{
&F(\alpha,\beta;\gamma;1-x)=
{\Gamma(\gamma)\Gamma(\gamma-\beta-\alpha)\over
\Gamma(\gamma-\alpha)\Gamma(\gamma-\beta)}
F(\alpha,\beta;\alpha+\beta+1-\gamma;x)
+\cr
&x^{\gamma-\alpha-\beta}
{\Gamma(\gamma)\Gamma(\alpha+\beta-\gamma)\over
\Gamma(\alpha)\Gamma(\beta)}
F(\gamma-\alpha,\gamma-\beta;\gamma+1-\alpha-\beta;x)
\cr}}
The hypergeometric function in two variables $F_1(x,y)$ can be defined
as the analytic continuation of the small $x,y$ ($|x|,|y|<1$)
expansion:
\eqn\foneex{F_1(\alpha,\beta,\beta';\gamma;x,y)=
\sum_{m=0}^{\infty}\sum_{n=0}^{\infty}
{(\alpha)_{m+n}(\beta)_m(\beta')_n\over m!n!(\gamma)_{m+n}}x^m y^n~,}
where $(a)_n$ is defined in \AACA.
Some useful identities are:
\eqn\foneid{\eqalign{
&F_1(\alpha,\beta,\beta';\gamma;x,1)=
{\Gamma(\gamma)\Gamma(\gamma-\alpha-\beta')\over
\Gamma(\gamma-\alpha)\Gamma(\gamma-\beta')}
F(\alpha,\beta;\gamma-\beta';x)~, \cr
&F_1(\alpha,\beta,\beta';\gamma;x,x)=F(\alpha,\beta+\beta';\gamma;x)~.\cr
}}
More useful identities:
\eqn\AAG{\Gamma(x)\Gamma(1-x)={\pi\over\sin(\pi x)}}
\eqn\AAH{\Gamma(1+ix)\Gamma(1-ix)={\pi x\over\sinh(\pi x)}, \qquad x\in R}
\eqn\AAI{\sin x + \sin y = 2\sin{x+y\over 2}\cos{x-y\over 2}}
\eqn\AAJ{\sinh x + \sinh y = 2\sinh{x+y\over 2}\cosh{x-y\over 2}}

\appendix{B}{The limit $z\to\infty$, $x\to\infty$ of $F_A$, $F_B$, $F_C$}

In section 3 we used the behavior of certain combinations
of hypergeometric functions in two variables, as their
arguments $x,z$ go to infinity. In this appendix we outline
the derivation of the result, eq. \behinf.

In order to study the limit $z\to\infty$, $x\to\infty$ of the
hypergeometric function $F_1(\alpha,\beta,\beta';\gamma;x,z)$
it is convenient to study its transformation under $z\to1/z$,
$x\to 1/x$. In \AAKK\ there is a function, denoted by $Z_3$,
that implements precisely this transformation:
\eqn\zthree{Z_3=x^{-\beta}z^{-\beta'}F_1(\beta+\beta'+1-\gamma,
\beta,\beta';\beta+\beta'+1-\alpha;{1\over x}, {1\over z}).}
Thus to study the behavior of $F_A$ \fabc\ as $z,x\to\infty$
it would be convenient to express $Z_3$ in terms of $F_A$,
$F_B$ and $F_C$, or alternatively in terms of $Z_1$, $Z_5$, and
$Z_8$ (see \AAKK\ for the general definition of $Z_5$; for our values
of the parameters it is equal to $F_B$).

In fact, as explained in \AAKK, precisely such a relation does indeed
exist. One has
\eqn\zthzzz{Z_3=c_1Z_1+c_5Z_5+c_8Z_8}
where $c_i$, $i=1,5,8$ are constants (independent of $x,z$).
Similarly, in order to study the large $z,x$ behavior of
$Z_5$ and $Z_8$ one is interested in the functions $Z_9$ and $Z_4$
in \AAKK, and again, they can be expressed in terms of the
basic functions as
\eqn\zninefour{\eqalign{
Z_9=&d_1Z_1+d_5Z_5+d_8Z_8\cr
Z_4=&e_1Z_1+e_5Z_5+e_8Z_8\cr
}}
If one knows the coefficients $c_i$, $d_i$, $e_i$, one can
analyze the behavior of the blocks $F_A$, $F_B$, $F_C$ in the
limit $z,x\to\infty$.

Determining all the coefficients is somewhat tedious,
but happily the part that is needed for deriving \behinf\ is rather simple.
First recall that we are only interested in the constant terms in
$F_A$, $F_B$ and $F_C$ in the limit $z,x\to\infty$. These correspond
to the contribution of the identity block in the limit studied in the text.
Terms that depend on $x,z$ in this limit give the contributions of the
boundary blocks $\Psi_t$ and $\Psi_{1-t}$ (which, as we mention in the
text and discuss below, turn out to vanish in the end).

One can show by studying the small
$x,z$ behavior of $Z_1$, $Z_5$, $Z_8$, that only the $Z_5$ terms on
the r.h.s. of \zthzzz, \zninefour\ contribute to the identity block; the
other terms contribute to the blocks corresponding to $\Psi_t$ and
$\Psi_{1-t}$. Therefore, we need only to compute $c_5$, $d_5$ and $e_5$
to derive \behinf.

The second nice fact is that if we set $x=1$ in \zthzzz, \zninefour,
the hypergeometric functions in two variables can be expressed in terms
of standard hypergeometric functions; $Z_1$ and $Z_8$ give rise to the
same hypergeometric function, while $Z_5$ produces a different one.
Thus, by setting $x=1$ one can compute precisely the coefficients
that are needed for \behinf. By using standard relations between
the hypergeometric functions $F$ and $F_1$, one finds
\eqn\cdefive{
\eqalign{
c_5=&{\Gamma(\beta+\beta'+1-\alpha)\Gamma(\gamma-\beta-1)
\over\Gamma(\beta')\Gamma(\gamma-\alpha)}(-)^{\beta+\beta'+1-\gamma}\cr
d_5=&{\Gamma(\alpha+1-\beta')\Gamma(\gamma-\beta-1)
\over\Gamma(\alpha)\Gamma(\gamma-\beta-\beta')}(-)^{\alpha+\beta+1-\gamma}\cr
e_5=&{\Gamma(\beta'+2-\gamma)\Gamma(\gamma-\beta-1)
\over\Gamma(\beta')\Gamma(1-\beta)}(-)^{\beta+\beta'+1-\gamma}\cr
}}
Substituting the values of $\alpha,\beta,\beta',\gamma$ relevant for
our problem, which are
\eqn\abbg{\alpha=\beta=t,\;\beta'=2h+t-1,\;\gamma=2t}
leads to the results \behinf.

In section 3 we argued that when the operator $\Phi_{t\over 2}$
approaches the boundary of the worldsheet ($z\to\bar z$) it only creates
the identity operator (\ie\ the coefficient of $\Psi_t$ and $\Psi_{1-t}$,
which also seem to be allowed by the OPE \ooe, vanish).
To prove this one notes that the function $\CF$ in \diskconfbl\
can be expanded in terms of the blocks relevant for the
$z\to\bar z$ limit, $Z_3,Z_4,Z_9$ given in eqs. \zthree\ -- \zninefour,
as follows:
\eqn\thfoni{\CF=a_3Z_3+a_4Z_4+a_9Z_9}
The first two terms on the r.h.s. are linear combinations of the $\Psi_t$
and $\Psi_{1-t}$ blocks, while $a_9$ is the
coefficient of the block of the identity.
Plugging \aabbccdd\ into \diskconfbl\ one finds that $a_3=a_4=0$.
Therefore, the amplitude for $\Phi_{t\over 2}$ to create
$\Psi_t$ and $\Psi_{1-t}$ on the boundary vanishes.

\listrefs
\end